\documentclass[useAMS,usenatbib]{mn2e}
\usepackage[dvips]{graphicx}
\newcommand{\bz}{$\langle B_z \rangle$}
\newcommand{\nz}{$\langle N_z \rangle$}
\newcommand{\vsini}{$v \sin i$}
\newcommand{\kms}{km\,s$^{-1}$}

\newcommand{\teff}{$T_{\rm eff}$}
\newcommand{\ra}{$R_{\rm A}$}
\newcommand{\rk}{$R_{\rm K}$}

\title[The tidally disrupted magnetosphere of HD 156324]{HD 156324: a tidally locked magnetic triple spectroscopic binary with a disrupted magnetosphere}
\author[M. Shultz]
{M.\ Shultz$^{1}$\thanks{E-mail: matthew.shultz@physics.uu.se},
Th.\ Rivinius$^{2}$,
G.\ A. Wade$^{3}$,
E.\ Alecian$^{4,5,6}$,
V.\ Petit$^{7}$
\newauthor{and the BinaMIcS Collaboration} \\
$^1$Department of Physics and Astronomy, Uppsala University, Box 516, Uppsala 75120 \\
$^2$ESO - European Organisation for Astronomical Research in the Southern Hemisphere, Casilla 19001, Santiago 19, Chile\\
$^3$Department of Physics, Royal Military College of Canada, Kingston, Ontario K7K 7B4, Canada\\
$^4$Universit\'e Grenoble Alpes, IPAG, F-38000 Grenoble, France\\
$^5$CNRS, IPAG, F-38000 Grenoble, France\\
$^6$LESIA, Observatoire de Paris, CNRS UMR 8109, UPMC, Universit\'e Paris Diderot, 5 place Jules Janssen, 92190, Meudon, France\\
$^7$Department of Physics and Astronomy, University of Delaware, 217 Sharp Lab, Newark, Delaware, 19716, USA\\
}
\begin{document}

\date{}

\pagerange{\pageref{firstpage}--\pageref{lastpage}} \pubyear{2002}

\maketitle

\label{firstpage}

\begin{abstract}
HD 156324 is an SB3 (B2V/B5V/B5V) system in the Sco OB4 association. The He-strong primary possesses both a strong magnetic field, and H$\alpha$ emission believed to originate in its Centrifugal Magnetosphere (CM). We analyse a large spectroscopic and high-resolution spectropolarimetric dataset. The radial velocities (RVs) indicate that the system is composed of two sub-systems, which we designate A and B. Period analysis of the RVs of the three components yields orbital periods $P_{\rm orb} = 1.5806(1)$~d for the Aa and Ab components, and 6.67(2)~d for the B component, a PGa star. Period analysis of the longitudinal magnetic field \bz~and H$\alpha$ equivalent widths, which should both be sensitive to the rotational period $P_{\rm rot}$ of the magnetic Aa component, both yield $\sim$1.58~d. Since $P_{\rm orb} = P_{\rm rot}$ Aa and Ab must be tidally locked. Consistent with this, the orbit is circularized, and the rotational and orbital inclinations are identical within uncertainty, as are the semi-major axis and the Kepler corotation radius. The star's H$\alpha$ emission morphology differs markedly from both theoretical and observational expectations in that there is only one, rather than two, emission peaks. We propose that this unusual morphology may be a consequence of modification of the gravitocentrifugal potential by the presence of the close stellar companion. We also obtain upper limits on the magnetic dipole strength $B_{\rm d}$ for the Ab and B components, respectively finding $B_{\rm d} < 2.6$~kG and $<0.7$~kG. 
\end{abstract}


\section{Introduction}

Early-type stars with magnetic fields are inherently rare, with only $\sim 7$\% of such stars possessing detectable photospheric magnetic fields \citep{grun2012c,2016MNRAS.456....2W,2017MNRAS.465.2432G}. When present, these fields tend to be strong (on the order of 1 kG), organized (predominantly dipolar), and stable on a timescale at least of decades. These properties have led to their characterization as fossil magnetic fields, i.e.\ remnants of a magnetic field produced or accumulated during a previous period in the star's evolution, in contrast to the magnetic fields of cool stars (i.e.\ stars with convective envelopes), which are continuously generated via dynamo action. 

Approximately 25\% of magnetic B-type stars show H$\alpha$ emission \citep{petit2013,my_phd_thesis}, which is believed to originate in the stars' Centrifugal Magnetospheres (CMs). A CM forms due to magnetic confinement of a star's ionized wind, together with centrifugal support due to rapid stellar rotation and forced corotation of the magnetically confined plasma, which typically prevents gravitational infall of material confined more than 1-2 stellar radii above the surface. The basic observed properties of CM H$\alpha$ emission morphology and variability are well-matched by the Rigidly Rotating Magnetosphere model (RRM; \citealt{town2005c}). RRM predicts two emission peaks with large velocity extrema ($\pm 2-3$\vsini) arising due to magnetospheric clouds located at the intersections of the magnetic and rotational equators. Over the course of a stellar rotational cycle, the emission peaks travel between the velocity extrema in an approximately sinusoidal fashion due to the changing projection on the sky of the magnetosphere. This basic morphology and pattern of variability has been observed in over a dozen rapidly rotating magnetic early B-type stars (e.g.\ $\sigma$ Ori E, \citealt{walborn1974,town2005b,2015MNRAS.451.2015O}; $\delta$~Ori~C, \citealt{leone2010}; HD 176582, \citealt{bohl2011}; HR 5907, \citealt{grun2012}; HR 7355, \citealt{rivi2013}; ALS 3694, \citealt{2014arXiv1411.2534S}). 

While magnetic fields are rare in hot stars, and magnetic hot stars with H$\alpha$ emission even rarer, close binaries containing a magnetic early-type star are rarer still: only 5 such systems with orbital periods less than 1 month are known, and the Binarity and Magnetic Interactions in various classes of Stars (BinaMIcS) Large Programs' (LP) Survey Component (SC) failed to detect a single new such system, establishing an upper limit of $\le 2$\% for their incidence \citep{2015IAUS..307..330A}.  

\cite{alecian2014} reported HD 156324 to be one of the very few systems that combines all three properties of magnetism, H$\alpha$ emission, and binarity. The system is an SB3 in the Sco OB4 association \citep{2005AA...438.1163K}, with a He-strong B2V star and a chemically normal B5V star with correlated, short-term radial velocity variability, indicating that they are physically associated in a close orbit. The third component was identified by \cite{alecian2014} as a PGa star, a class of chemically peculiar stars proposed to be a hotter extension of the non-magnetic HgMn stars \citep{2006ARep...50..123R,2014MNRAS.442.3604H}. It also exhibits moderate RV variation, and its relationship to the first two stars is unclear. We designate the components Aa, Ab, and B, respectively\footnote{\cite{alecian2014} designated them A, B, and C.}. The magnetic field is associated with the Aa component, and the H$\alpha$ emission is also thought to belong to this star.

Due to the small number of magnetic and spectroscopic measurements available, \cite{alecian2014} were unable to determine rotational or orbital ephemerides. In this paper we report the results of systematic high-resolution spectropolarimetric and spectroscopic observation campaigns aimed at addressing this problem. The observations are described in \S~\ref{sec:obs}. Radial velocity (RV) measurements and orbital periods are presented in \S~\ref{sec:rv}. \S~\ref{sec:mag} describes the magnetometry, and the H$\alpha$ variability is examined in \S~\ref{sec:halpha}. Based on the results presented in \S~\ref{sec:mag} and \ref{sec:halpha}, we are able to derive the rotational period. The orbital, stellar, magnetic, rotational, and magnetospheric parameters are derived and discussed in \S~\ref{sec:discussion}, and the conclusions summarized in \S~\ref{sec:conclusions}.

\section{Observations}\label{sec:obs}

\subsection{ESPaDOnS spectropolarimetry}

ESPaDOnS is a fibre-fed echelle spectropolarimeter at the Canada-France-Hawaii Telescope (CFHT). It has a spectral resolution $\lambda/\Delta\lambda \sim 65,000$, and a spectral range from 370 to 1050 nm over 40 spectral orders. Each observation consists of 4 polarimetric sub-exposures, between which the orientation of the instrument's Fresnel rhombs are changed, yielding 4 intensity (Stokes $I$) spectra, 1 circularly polarized (Stokes $V$) spectrum, and 2 null polarization ($N$) spectra, the latter obtained in such a way as to cancel out the intrinsic polarization of the source and to evaluate spurious polarization contributions. \cite{2016MNRAS.456....2W} describe the reduction and analysis of ESPaDOnS data in detail. Nine new Stokes $V$ observations were acquired between 04/2014 and 06/2014 by a P.I. (Principal Investigator) program\footnote{Program Code CFHT 14AC010, P.I. M. Shultz.}. A further 12 observations were collected between 07/2015 and 08/2015 under the auspices of the BinaMIcS Large Program. A uniform sub-exposure time of 450 s was used for all observations. The median peak signal-to-noise (S/N) per spectral pixel is 369. 

\subsection{HARPSpol spectropolarimetry}

HARPSpol is a high-resolution ($\lambda/\Delta\lambda\sim 100,000$) echelle spectropolarimeter with a spectral range covering 378--691 nm, with a gap between 524 and 536 nm, across 71 spectral orders. It is installed at the 3.6~m telescope at the European Southern Observatory (ESO) La Silla facility. As with ESPaDOnS, each spectropolarimetric sequence consists of 4 polarized sub-exposures, which are combined to yield the Stokes $V$ spectrum as well as a diagnostic null $N$. The sub-exposure time was 900~s. Two observations were acquired in 2012 by the Magnetism in Massive Stars (MiMeS) ESO Large Program. The acquisition, reduction, and analysis of these data were described by \cite{alecian2014}. An additional HARPSpol observation was acquired on 03/07/2015 by an independent observing program\footnote{Program code 095.D-0269(A), P.I.\ C.\ Neiner.}. This observation was reduced in the same fashion as the MiMeS data.

\subsection{FEROS spectroscopy}

FEROS is a high-dispersion echelle spectrograph, with $\lambda/\Delta\lambda\sim$48,000 and a spectral range of 375--890 nm \citep{1998SPIE.3355..844K}. It is mounted at the 2.2~m La Silla MPG telescope. We acquired 11 spectra between 06/2015 and 07/2015, with an exposure time of 1400 s\footnote{Program code MPIA LSO22-P95-007, P.I. M. Shultz.}. The data were reduced using the standard FEROS Data Reduction System MIDAS scripts\footnote{Available at https://www.eso.org/sci/facilities/lasilla/\\instruments/feros/tools/DRS.html}. The median peak S/N per spectral pixel is 261.

\section{Radial velocities}\label{sec:rv}

\begin{figure}
\centering
\includegraphics[width=\hsize]{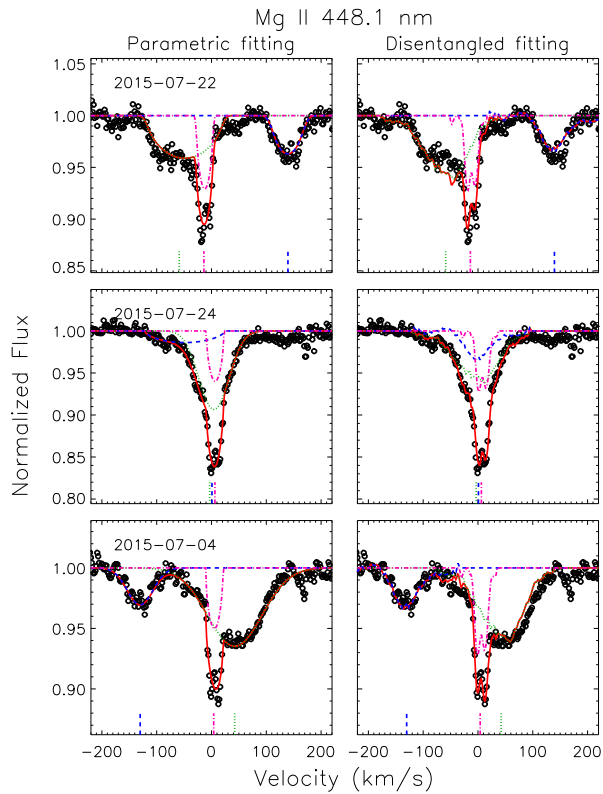}
\caption{Radial velocity measurements using the Mg~{\sc ii}~448.1 nm line at the maximum and minimum RV separation of the Aab components. {\em Left}: multi-component line-profile fits. Observations are shown by black circles. The dotted (dark green) curve shows the fit to Aa; the dashed (blue) curve shows Ab; the dot-dashed (purple) curve shows the B component; the solid (red) curve shows the full profile. The RVs of the three components are indicated by short lines at the bottom of each panel, with colours and line style corresponding to the model fits. {\em Right}: as on the left, but using the mean disentangled profiles. Note the strong blending the Aab pair on 2015-07-24, and that disentangling more faithfully reproduces the line profile of the B component.}
\label{MgII4481_rv}
\end{figure}

Radial velocities were measured using the Mg~{\sc ii} 448.1 nm line, in which all three components are clearly visible, using the {\sc idl} program {\sc fit\textunderscore lsd\textunderscore binary}, described by \cite{2017MNRAS.465.2432G}. Representative fits are shown in the left panels of Fig.\ \ref{MgII4481_rv} for the observations with the maximum and minimum RV separation of the Aab components, at the maximum (top) and minimum (bottom) RV of the Aa component. 

This step of the analysis also yielded line broadening measurements, which we paramaterize using the projected rotational velocity \vsini~and the (radial-tangential) macroturbulent velocity $v_{\rm mac}$. We obtained: \vsini$_{\rm Aa} = 53 \pm 10$~\kms, $v_{\rm mac,Aa} = 44 \pm 29$~\kms; \vsini$_{\rm Ab} = 32 \pm 10$~\kms, $v_{\rm mac,Ab} = 31 \pm 25$~\kms; and \vsini$_{\rm B} = 15 \pm 3$~\kms, $v_{\rm mac,B} < 10$~\kms. The velocity fields and their uncertainties were obtained from observations at which the components were separated, and computed from the median and standard deviation of the fits across all such observations. While the B component is always blended, its very sharp lines mean that the FWHM of its line profile is easily distinguished in all spectra. Close inspection of the B component's Mg~{\sc ii} 448.1 nm line, however, shows an atypical profile that may indicate the presence of two partially blended lines. We therefore searched the spectrum for unblended lines dominated by the B component, and identified Fe~{\sc ii} 516.9 nm, P~{\sc ii} 525.3 nm and P~{\sc ii} 545.6 nm as the strongest such lines. These yielded \vsini$_{\rm B} = 5 \pm 2$~\kms~and $v_{\rm mac,B} = 9 \pm 3$~\kms. While these lines all yielded lower \vsini~than obtained via Mg~{\sc ii} 448.1 nm, there was no difference in RV. 

When Aa and Ab are strongly blended, unique solutions consistent with the line broadening parameters obtained when the components are clearly separated could not be derived, making the RVs difficult to determine precisely. Therefore the RVs were refined by disentangling the profiles using an iterative algorithm as described by \cite{2006AA...448..283G}, with the RVs re-calculated at each iteration according to the centre of gravity of the residual profiles, as described by \cite{2017MNRAS.465.2517W}. Individual and composite fits using the mean disentangled profiles of the three components are shown in the right-hand panels of Fig.\ \ref{MgII4481_rv}. The RVs obtained in this fashion (first via parametric fitting, and then refined via iterative disentangling) are listed in Table \ref{rvewtab}.

\begin{table}
\centering
\caption{Log of radial velocity (RV) and H$\alpha$ EW measurements. Instrument refers to HARPSpol (H), ESPaDOnS (E), or FEROS (F). Estimated RV uncertainties are 10~\kms~for the Aab components, and 5~\kms~for the B component.}
\label{rvewtab}
\resizebox{8.5 cm}{!}{
\begin{tabular}{l l l r r r r }
\hline
\hline
\\
          &      &       & \multicolumn{3}{c}{RV~(\kms)} & \\
HJD       & Calendar & Inst. & Aa & Ab & B & H$\alpha$ EW \\
$-2450000$  & Date &       &   &   &  & (nm)         \\
\hline
6127.62863 & 19/07/2012 &  H &   5 &   62 & -13 & 0.145$\pm$0.011 \\
6128.78131 & 20/07/2012 &  H & -41 &  145 &  -1 & 0.119$\pm$0.016 \\
6766.05747 & 18/04/2014 &  E & -39 &   95 &  27 & 0.128$\pm$0.008 \\
6770.04427 & 22/04/2014 &  E &  42 &  -71 &  29 & 0.175$\pm$0.010 \\
6770.06746 & 22/04/2014 &  E &  38 &  -62 &  30 & 0.167$\pm$0.010 \\
6770.09045 & 22/04/2014 &  E &  34 &  -53 &  30 & 0.169$\pm$0.009 \\
6816.95336 & 08/06/2014 &  E &  29 &  -55 &  23 & 0.168$\pm$0.005 \\
6817.95616 & 09/06/2014 &  E & -46 &  139 &  12 & 0.116$\pm$0.005 \\
6817.97947 & 09/06/2014 &  E & -48 &  141 &  12 & 0.114$\pm$0.005 \\
6818.95181 & 10/06/2014 &  E &  49 & -103 &  17 & 0.179$\pm$0.005 \\
6818.97457 & 10/06/2014 &  E &  49 &  -90 &  17 & 0.178$\pm$0.005 \\
7203.61047 & 30/06/2015 &  F & -40 &  137 &  -2 & 0.122$\pm$0.011 \\
7203.67273 & 30/06/2015 &  F & -51 &  140 &  -2 & 0.116$\pm$0.011 \\
7204.65278 & 01/07/2015 &  F &  48 &  -91 & -18 & 0.177$\pm$0.010 \\
7204.71380 & 01/07/2015 &  F &  43 &  -67 & -18 & 0.171$\pm$0.010 \\
7205.47959 & 01/07/2015 &  F & -35 &   81 & -18 & 0.131$\pm$0.011 \\
7205.64923 & 02/07/2015 &  F & -17 &    0 & -16 & 0.148$\pm$0.011 \\
7206.48243 & 02/07/2015 &  F &  12 &   20 &  -8 & 0.155$\pm$0.013 \\
7206.61774 & 03/07/2015 &  H & -28 &   97 &  -4 & 0.132$\pm$0.016 \\
7206.68900 & 03/07/2015 &  F & -42 &  124 &  -6 & 0.121$\pm$0.011 \\
7207.53483 & 04/07/2015 &  F &  42 & -119 &   0 & 0.194$\pm$0.011 \\
7207.69888 & 04/07/2015 &  F &  42 & -130 &   4 & 0.188$\pm$0.011 \\
7208.53337 & 05/07/2015 &  F & -61 &  116 &   5 & 0.138$\pm$0.013 \\
7225.82139 & 22/07/2015 &  E & -58 &  139 &   6 & 0.115$\pm$0.005 \\
7226.80460 & 23/07/2015 &  E &  37 &  -85 &  10 & 0.173$\pm$0.005 \\
7227.78663 & 24/07/2015 &  E &  -3 &    0 &   9 & 0.147$\pm$0.005 \\
7228.92234 & 25/07/2015 &  E & -39 &  138 &  -4 & 0.120$\pm$0.005 \\
7229.84399 & 26/07/2015 &  E &  47 & -128 & -15 & 0.185$\pm$0.005 \\
7230.75400 & 27/07/2015 &  E & -37 &   94 &  -8 & 0.121$\pm$0.005 \\
7231.83172 & 28/07/2015 &  E &  -4 &   79 &  -8 & 0.138$\pm$0.005 \\
7232.77609 & 29/07/2015 &  E &  30 & -103 &   0 & 0.177$\pm$0.017 \\
7232.79952 & 29/07/2015 &  E &  41 & -108 &  10 & 0.176$\pm$0.008 \\
7233.76335 & 30/07/2015 &  E & -42 &  137 &   3 & 0.114$\pm$0.005 \\
7235.83683 & 01/08/2015 &  E &  25 &  -70 &  15 & 0.172$\pm$0.005 \\
7236.84352 & 02/08/2015 &  E & -46 &  128 &  11 & 0.115$\pm$0.005 \\
\hline
\hline
\end{tabular}
}
\end{table}

\begin{figure}
\centering
\includegraphics[width=\hsize]{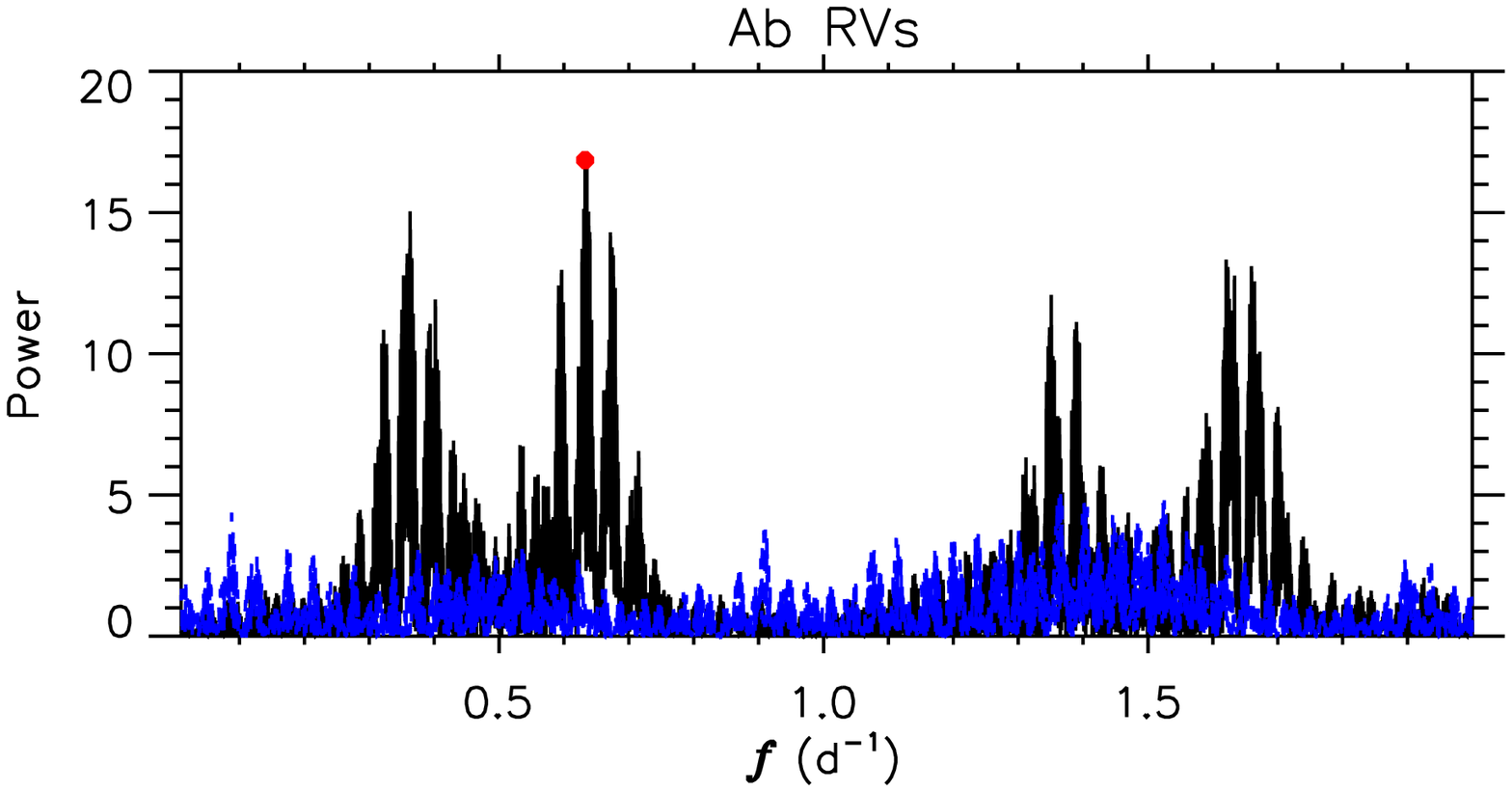}
\includegraphics[width=\hsize]{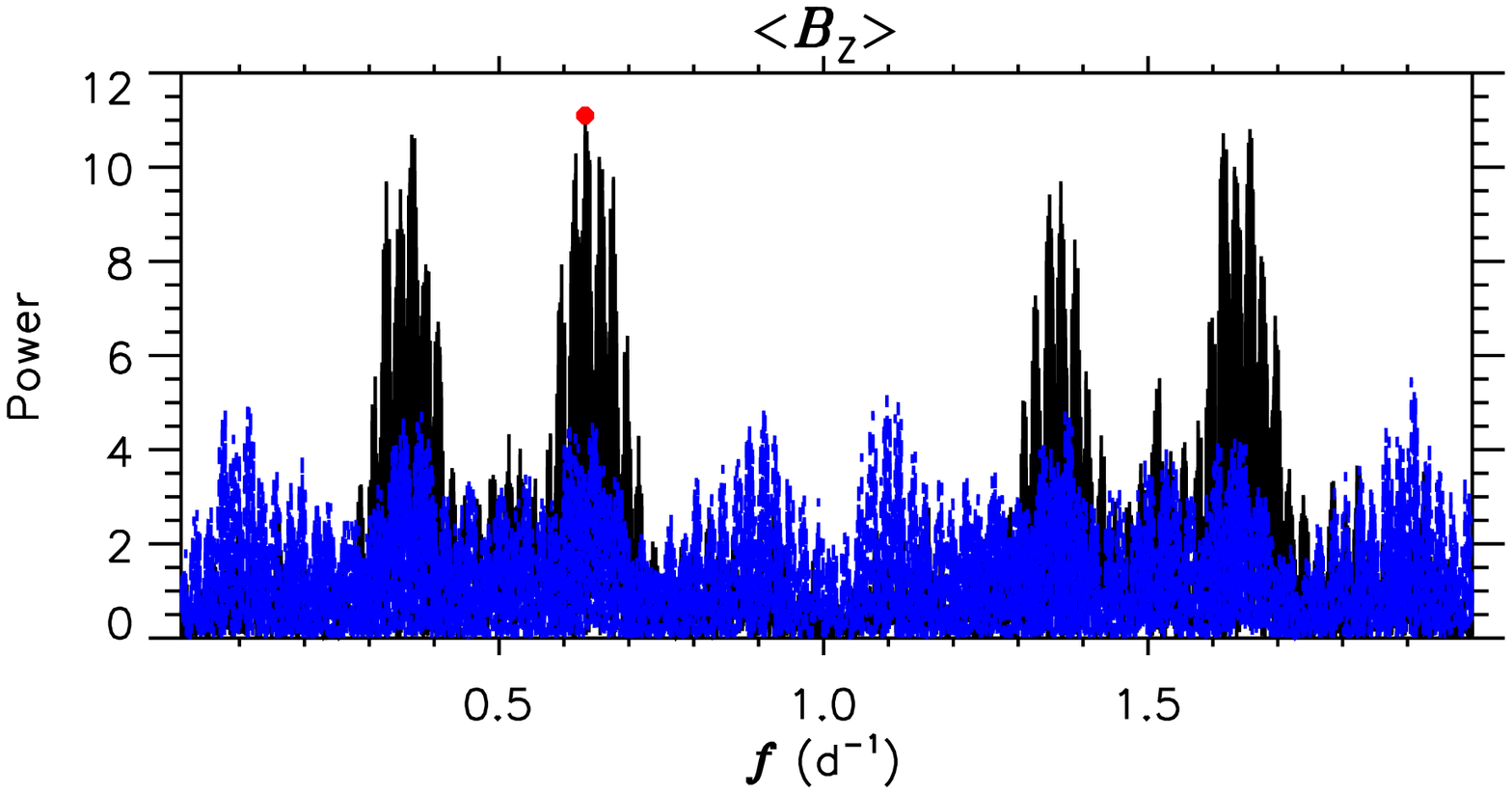}
\includegraphics[width=\hsize]{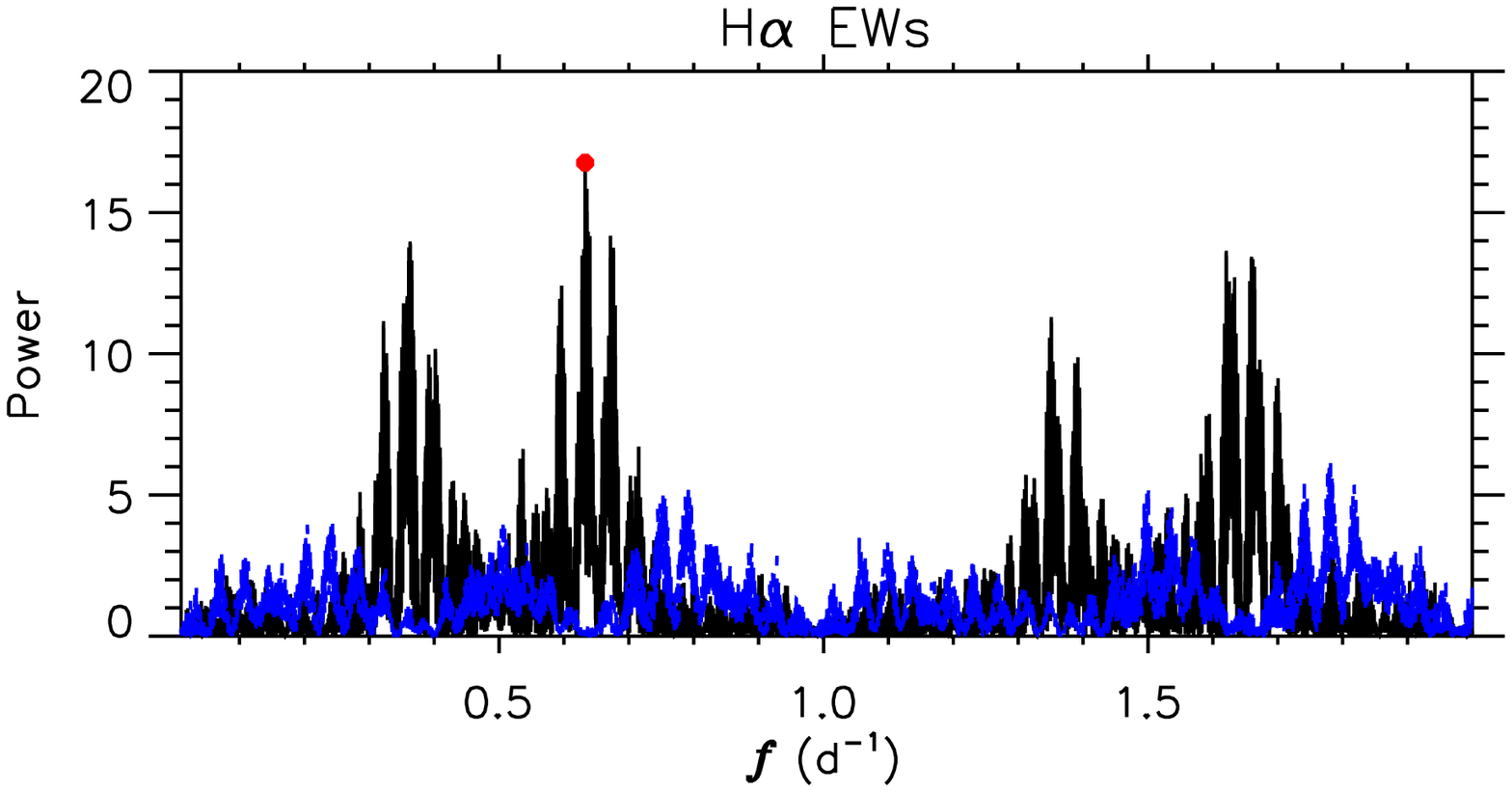}
\caption{{\em Top}: Frequency spectrum for the Ab component's RVs (solid black), and for synthetic noise with the same time sampling as the measurements (dashed blue). The red circle indicates the maximum amplitude period at $f=0.63260(4)~{\rm d}^{-1}$ (1.5806(1)~d). {\em Middle}: as above, for \bz~(solid black) and \nz~{dashed blue}. Maximum power is at $f = 0.6327(1)~{\rm d}^{-1}$ (1.5804(3)~d). {\em Bottom}: as above, for H$\alpha$ EWs. Maximum power is at $f = 0.6326(1)~{\rm d}^{-1}$ (1.5806(3)~d). Maximum power occurs at the same period is the same in all 3 periodograms. The two other highest peaks occur at the aliases $|f-1|$ and $f+1$.}
\label{periods}
\end{figure}

\begin{figure}
\centering
\includegraphics[width=9cm]{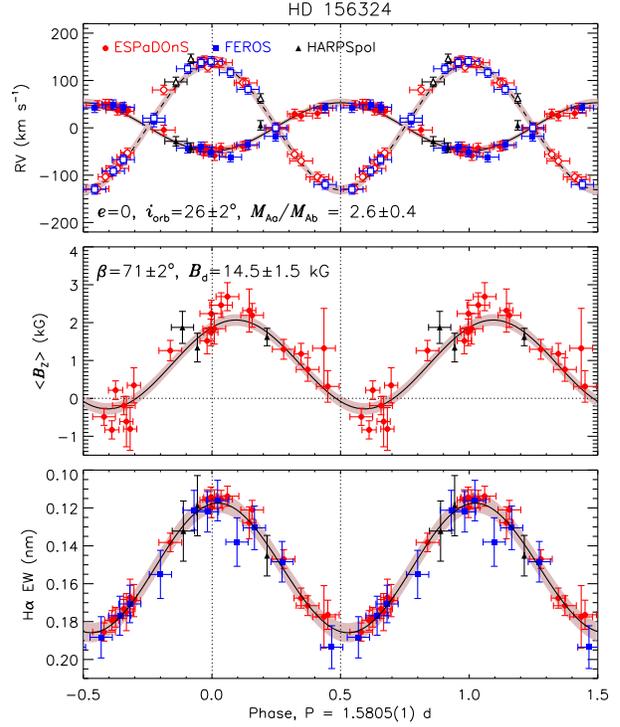}
\caption{Radial velocities ({\em top}), \bz~({\em middle}), and H$\alpha$ EWs ({\em bottom}). All three measurements show a coherent variation with the same period. In the top panel, Aa RVs are indicated with filled symbols, Ab with open symbols. The solid and dashed curves in the top panel are synthetic RV curves corresponding to a Keplerian orbit for Aa and Ab respectively, using the measured RV semi-amplitudes and assuming an eccentricity $e=0$; shaded regions indicate 1$\sigma$ uncertainties in the orbital parameters. The solid curve and shaded region in the middle panel shows the least-squares sinusoidal fits and titsheir 1$\sigma$ uncertainties, used for modelling the surface magnetic field properties, and the dotted line indicates \bz~$=0$~G. Vertical dotted lines indicate phases 0.0 and 0.5: note the phase shift of about 0.1 cycles between the \bz~and RV extrema. The solid curve and shaded region in the bottom panel shows the 2$^{nd}$-order sinusoidal fit and its uncertainty to the H$\alpha$ EWs. There is no evidence for a EW double-wave variation, consistent with a star for which only one magnetic pole is visible during a rotational cycle.}
\label{bzrvew}
\end{figure}

\begin{figure}
\centering
\includegraphics[width=\hsize]{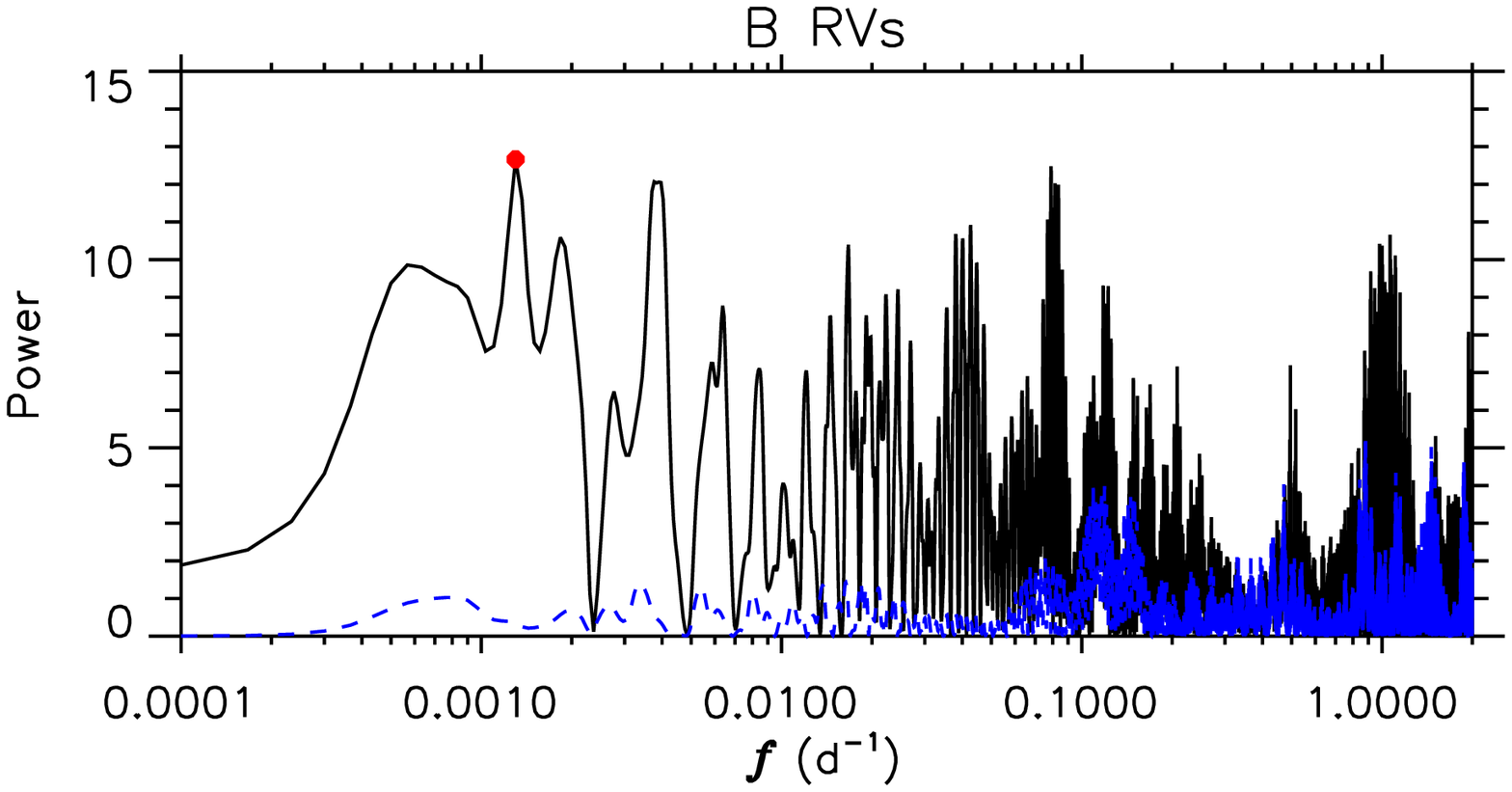}
\includegraphics[width=\hsize]{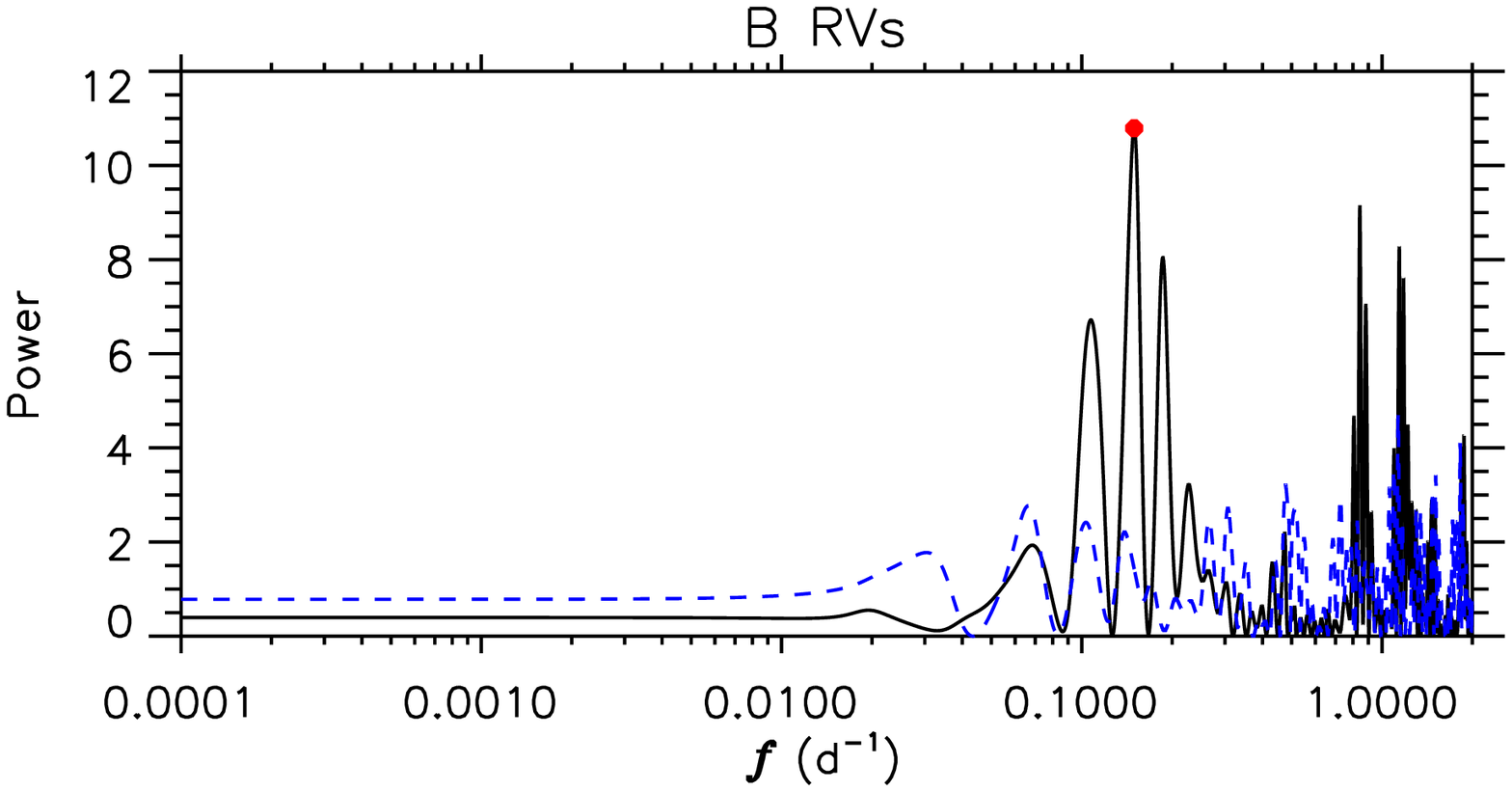}
\caption{{\em Top}: Frequency spectrum for the B component's RVs, using the full dataset (solid black), and synthetic noise with a standard deviation of 5 \kms~and an identical time sampling to the measurements. {\em Bottom}, as above, using only the 2015 data. The frequency spectra are shown on a log-scale in order to emphasize the low-frequency region. Maximum power in the full dataset occurs at $f=0.0013(1)~{\rm d}^{-1}$ (2.1(2) yr), while the 2015 dataset shows no power at low frequencies but strong power at $f = 0.149(6) {\rm d}^{-1}$ (6.7(3) d).}
\label{b_rv_periods}
\end{figure}

\begin{figure}
\centering
\includegraphics[width=\hsize]{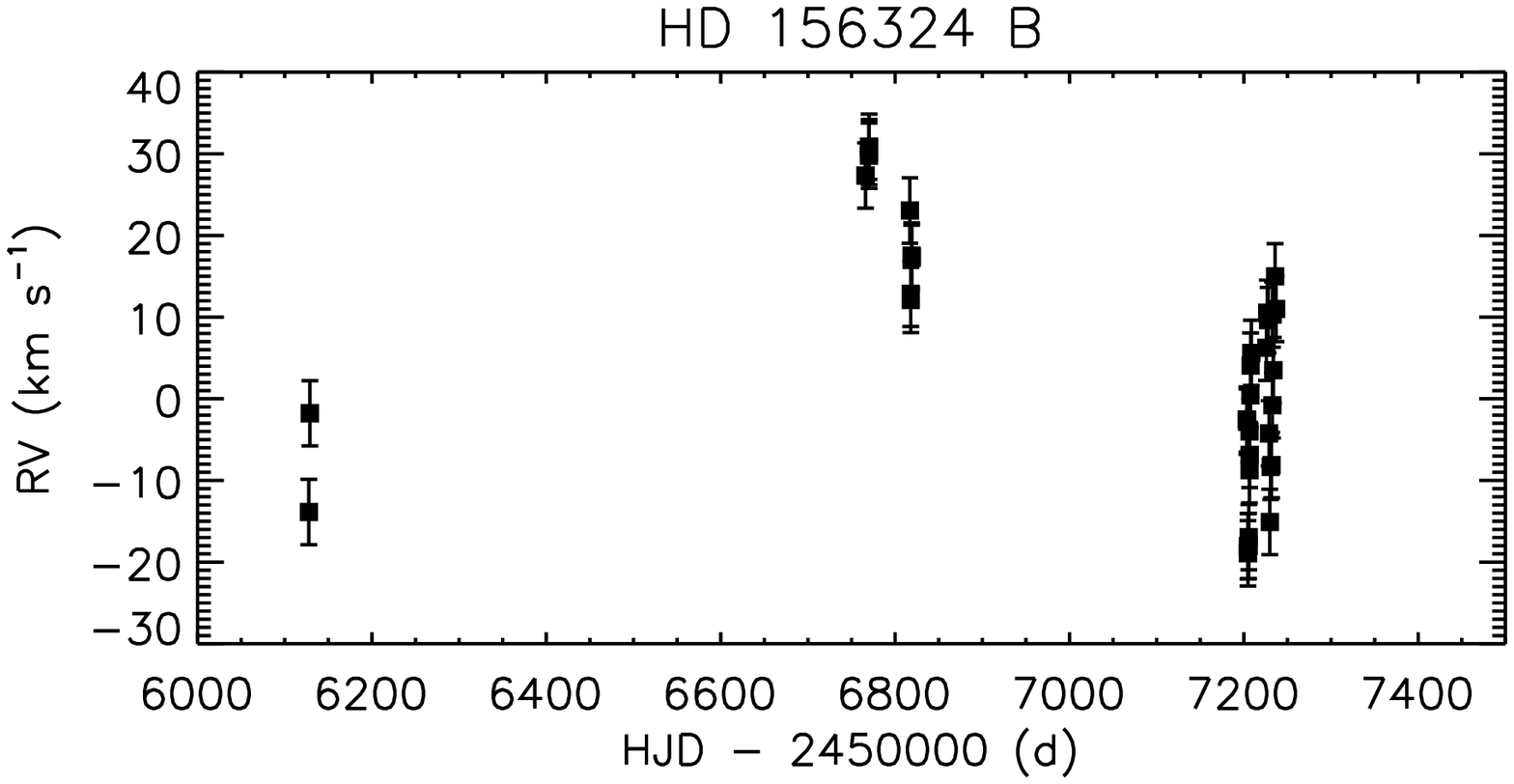}
\includegraphics[width=\hsize]{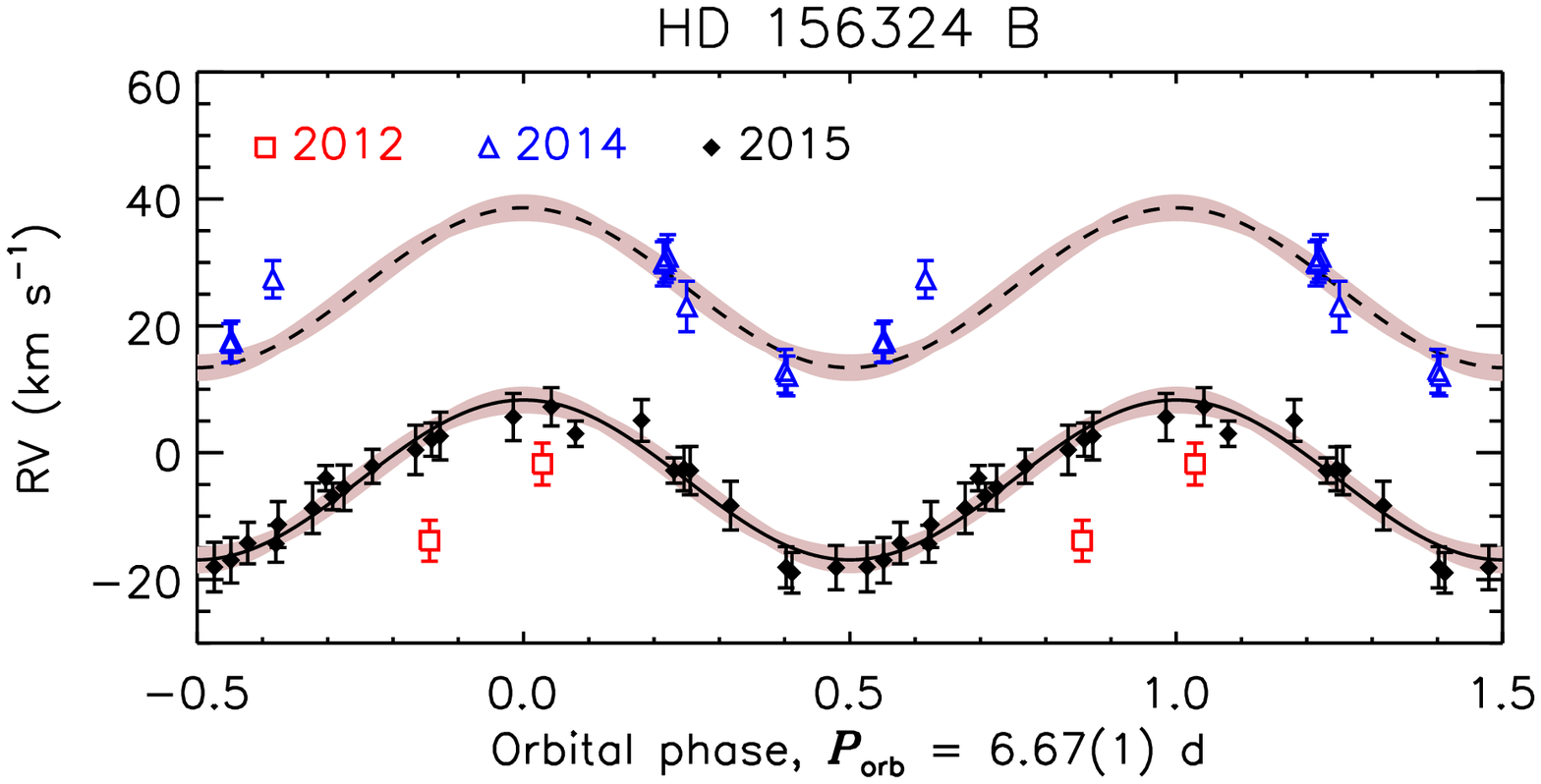}
\caption{Radial velocities for the B component, as a function of time (top), and phased with the period determined using the 2015 data (bottom). The solid curve shows the best-fit orbital model to the 2015 data; the dashed line is the same model, offset to $v_0 = 26$~\kms. Shaded regions show model uncertainties.}
\label{b_rv}
\end{figure}

A frequency search of the Ab component's RVs conducted using Lomb-Scargle statistics yielded maximum power at 0.63269(43)~d$^{-1}$, or 1.5805(10)~d, where the numbers in brackets indicate the uncertainty in the least significant digits. The frequency spectrum for the Ab component's RVs is shown in the top panel of Fig.\ \ref{periods}. The Aa component's RVs yield essentially identical results, albeit with an uncertainity of 0.00026~d due to the smaller amplitude. A period search conducted on synthetic noise measurements (gaussian noise with the same time-sampling and a standard deviation equal to the estimated uncertainty of the RVs, 10~\kms), with a standard deviation equivalent to the estimated uncertainty of the RVs, does not yield any significant power at 1.58~d. The False Alarm Probability (FAP) of this period is about $10^{-7}$, much lower than the minimum FAP of the synthetic noise periodogram of about 0.75. The S/N of the peak is 67; after pre-whitening, the maximum S/N in the periodogram is 3.3, below the threshold of 4 generally used as the minimum S/N for statistical significance in a periodogram \citep{1993A&A...271..482B,1997A&A...328..544K}. Given the high statistical significance of the peak, and the identical results for Aa and Ab, we identify this as the orbital period. The RVs of the Aab components are shown phased with this period in the top panel of Fig.\ \ref{bzrvew}, using the ephemeris

\begin{equation}\label{ephem}
T_{\rm max(RV)} ({\rm HJD}) = 2456127.29(2) + 1.5805(10) \times E.
\end{equation}

\noindent where $E$ is the HJD of the observation, ${\rm JD0} = 2456127.29(2)$ is defined by the maximum RV of the Ab component one cycle before the first observation in the dataset as determined via a least-squares sinusoidal fit, and the the uncertainty in JD0 was determined from the phase uncertainty of the sinusoidal fit. 

Conducting the same period analysis on the RVs of the B component yields maximum power at $0.0013(1)~{\rm d}^{-1}$, or 2.1(2) yr (Fig.\ \ref{b_rv_periods}, top). The top panel of Fig.\ \ref{b_rv} shows the B component's RVs as a function of time, and there does indeed appear to be a long-term modulation. Restricting the dataset to the measurements collected in 2015, the epoch with the greatest number of observations, yields a clear peak at 6.67(2)~d (Fig.\ \ref{b_rv_periods}, bottom). The B component's RVs are shown phased with this period in the bottom panel of Fig.\ \ref{b_rv}. While this period provides a coherent phasing of the 2015 data, the measurements collected in 2012 and 2014 are not coherently phased by this period. This is due to a systematic offset in the RVs between these epochs, with the 2014 RVs being about 30 \kms~higher than the 2015 RVs. Thus, the data suggest that the B component's RVs show two significant modulations, one on a timescale of days, the other on a timescale of years. 

\section{Magnetometry}\label{sec:mag}

\begin{table*}
\centering
\caption{Log of magnetic measurements. Instrument refers to either HARPSpol (H) or ESPaDOnS (E). Detection flags (DF) are definite detections (DD), marginal detections (MD), or non-detections (ND), according to the criteria given by \protect\cite{1992AA...265..669D,d1997}.}
\label{bztab}
\resizebox{18 cm}{!}{
\begin{tabular}{l r r r r r r r r r r r r r}
\hline
\hline
\\
          &       & \multicolumn{4}{c}{Aa} & \multicolumn{4}{c}{Ab} & \multicolumn{4}{c}{B} \\
HD        & Inst. & \bz & DF$_{V}$ & \nz & DF$_{N}$ & \bz & DF$_{V}$ & \nz & DF$_{N}$ & \bz & DF$_{V}$ & \nz & DF$_{N}$ \\
-2450000  &       & (G) &          & (G) &          & (G) &          & (G) &          & (G) &          & (G) &          \\
\hline
6127.62863 &  H & 1534$\pm$218 &  {\bf DD} &  124$\pm$217 &  {\bf DD} & -- & -- & -- & -- &   -19$\pm$ 134 &  ND &   -32$\pm$ 134 &  ND \\ 
6128.78131 &  H & 1297$\pm$327 &  {\bf DD} &  137$\pm$327 &  ND & -- & -- & -- & -- &   114$\pm$ 191 &  ND &  -236$\pm$ 192 &  ND \\ 
6766.05747 &  E & 2794$\pm$530 &  {\bf DD} &  307$\pm$527 &  ND & -- & -- & -- & -- &   -63$\pm$ 179 &  ND &   204$\pm$ 179 &  ND \\ 
6770.04427 &  E & -562$\pm$593 &  ND & 2090$\pm$594 &  ND & -- & -- & -- & -- &    42$\pm$ 125 &  ND &   111$\pm$ 125 &  ND \\ 
6770.06746 &  E & -742$\pm$494 &  MD &  -93$\pm$494 &  ND & -- & -- & -- & -- &   -23$\pm$  76 &  ND &     6$\pm$  76 &  ND \\ 
6770.09045 &  E &  287$\pm$419 &  {\bf DD} &   38$\pm$419 &  ND & -- & -- & -- & -- &   -63$\pm$  86 &  ND &   -73$\pm$  86 &  ND \\ 
6816.95336 &  E & 1051$\pm$248 &  {\bf DD} &  220$\pm$248 &  ND &-3685$\pm$2909 &  ND & -8079$\pm$3162 &  ND &     1$\pm$  72 &  ND &   -29$\pm$  72 &  ND \\ 
6817.95616 &  E & 1356$\pm$312 &  {\bf DD} &  103$\pm$312 &  ND &  682$\pm$1262 &  ND &  1764$\pm$1273 &  ND &    68$\pm$  79 &  ND &   -60$\pm$  79 &  ND \\ 
6817.97947 &  E & 1543$\pm$317 &  {\bf DD} & -206$\pm$316 &  ND & 1855$\pm$1318 &  ND &  -980$\pm$1305 &  ND &    -3$\pm$  71 &  ND &    -4$\pm$  71 &  ND \\ 
6818.95181 &  E & -696$\pm$207 &  ND &   14$\pm$207 &  ND & -978$\pm$1623 &  ND &  -394$\pm$1620 &  ND &    22$\pm$  65 &  ND &    85$\pm$  65 &  ND \\ 
6818.97457 &  E &  200$\pm$214 &  MD &  -19$\pm$214 &  ND & -- & -- & -- & -- &   -50$\pm$ 116 &  ND &   133$\pm$ 116 &  ND \\ 
7206.61774 &  H & 1693$\pm$381 &  {\bf DD} &   69$\pm$380 &  ND & -- & -- & -- & -- &   -23$\pm$  73 &  ND &   -82$\pm$  73 &  ND \\ 
7225.82139 &  E & 2077$\pm$278 &  {\bf DD} & -295$\pm$275 &  ND &  502$\pm$1392 &  ND &   944$\pm$1396 &  ND &     4$\pm$  70 &  ND &    55$\pm$  70 &  ND \\ 
7226.80460 &  E & -172$\pm$218 &  {\bf DD} &  -40$\pm$218 &  ND & 1391$\pm$2125 &  ND & -1243$\pm$2123 &  ND &   -82$\pm$  78 &  ND &    92$\pm$  78 &  ND \\ 
7227.78663 &  E & 1314$\pm$266 &  {\bf DD} &   49$\pm$265 &  ND & -- & -- & -- & -- &    26$\pm$  74 &  ND &    67$\pm$  74 &  ND \\ 
7228.92234 &  E & 1986$\pm$320 &  {\bf DD} & -226$\pm$318 &  ND &  879$\pm$1348 &  ND & -1342$\pm$1354 &  ND &    67$\pm$  74 &  ND &    26$\pm$  74 &  ND \\ 
7229.84399 &  E & -437$\pm$207 &  {\bf DD} & -167$\pm$207 &  ND &  122$\pm$ 974 &  ND &  -704$\pm$ 976 &  ND &    28$\pm$  74 &  ND &     1$\pm$  74 &  ND \\ 
7230.75400 &  E & 2060$\pm$295 &  {\bf DD} &  361$\pm$294 &  ND & -134$\pm$ 763 &  ND &   491$\pm$ 764 &  ND &    -8$\pm$  65 &  ND &   -62$\pm$  65 &  ND \\ 
7231.83172 &  E & 1177$\pm$246 &  {\bf DD} &  216$\pm$246 &  ND & -- & -- & -- & -- &   -23$\pm$ 307 &  ND &    14$\pm$ 307 &  ND \\ 
7232.77609 &  E & 1343$\pm$931 &  ND & 2393$\pm$932 &  ND & -- & -- & -- & -- &  -142$\pm$ 154 &  ND &   190$\pm$ 154 &  ND \\ 
7232.79952 &  E &  344$\pm$370 &  {\bf DD} &  465$\pm$370 &  ND & -- & -- & -- & -- &    30$\pm$  78 &  ND &   -30$\pm$  78 &  ND \\ 
7233.76335 &  E & 2399$\pm$330 &  {\bf DD} & -622$\pm$327 &  ND & -359$\pm$1638 &  ND &  2061$\pm$1660 &  ND &    52$\pm$  70 &  ND &   -20$\pm$  70 &  ND \\ 
7235.83683 &  E &  693$\pm$283 &  {\bf DD} &  179$\pm$283 &  ND & -- & -- & -- & -- &     5$\pm$  65 &  ND &   119$\pm$  65 &  ND \\ 
7236.84352 &  E & 1558$\pm$354 &  {\bf DD} &  281$\pm$353 &  ND & -632$\pm$1312 &  ND & -2188$\pm$1340 &  ND &   129$\pm$  97 &  ND &    77$\pm$  97 &  ND \\ 
\hline
\hline
\end{tabular}
}
\end{table*}

\subsection{Aa}

\begin{figure}
\centering
\includegraphics[width=\hsize]{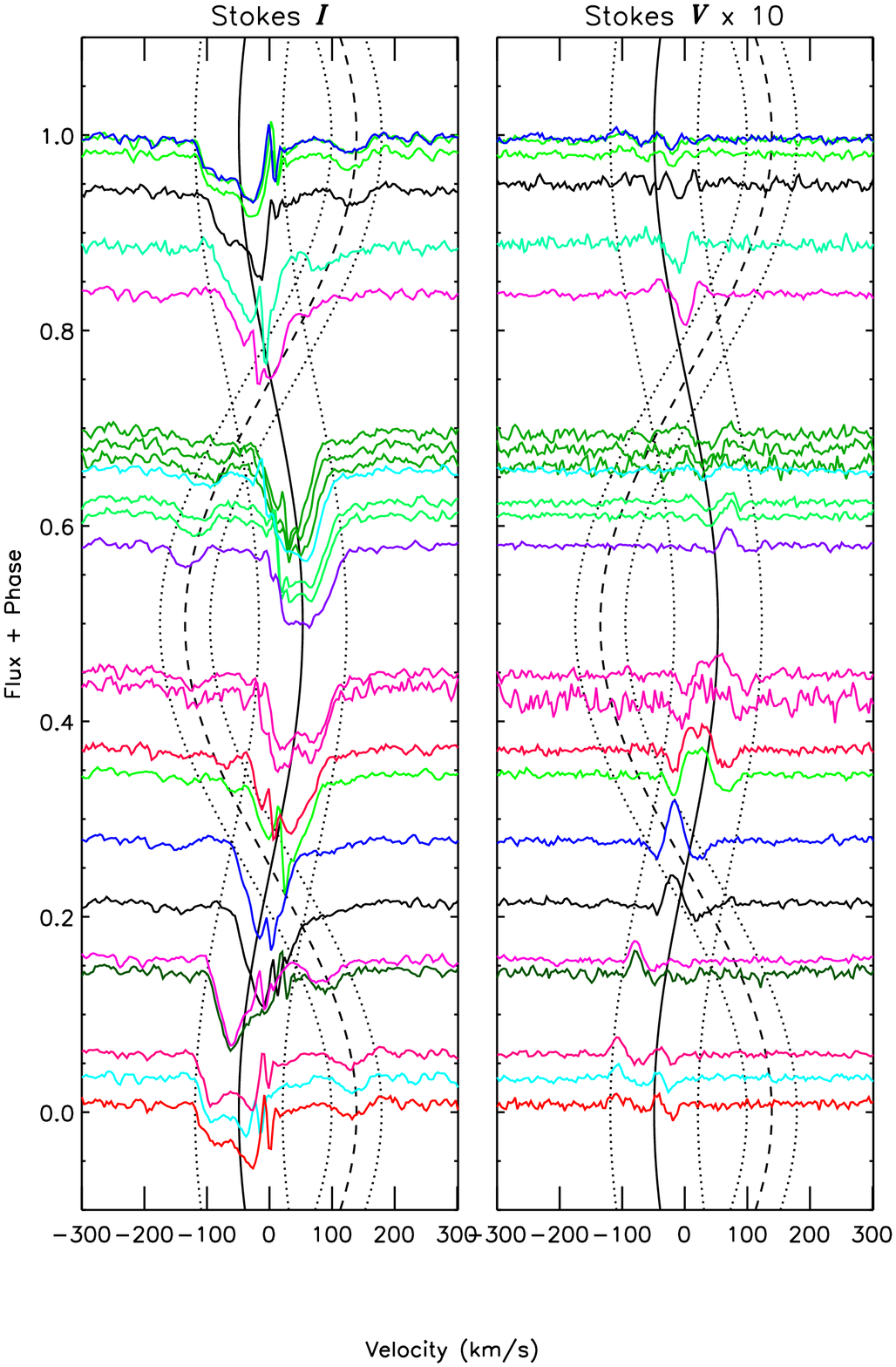}
\caption{LSD Stokes $I$ and $V$ profiles extracted with all metallic lines, arranged in order of orbital phase (Eqn.\ \ref{ephem}). The B component's contribution to the Stokes $I$ line profile has been removed by subtracting its mean disentangled profile. Different orbital/rotational cycles are indicated by different colours. The solid (dashed) curve shows the orbital model for Aa (Ab) from the top panel in Fig.\ \ref{bzrvew}; dotted curves show $\pm$\vsini~of the two components. The Stokes $V$ signature is clearly associated with HD 156324Aa. Furthermore, both Stokes $I$ and $V$ vary coherently with the orbital period.}
\label{lsd}
\end{figure}

Least-Squares Deconvolution (LSD) profiles were extracted from the spectropolarimetric data in order to achieve the highest possible S/N. Cleaning of the 22~kK line mask (see \S~\ref{subsec:stellarpars}) obtained from VALD3 \citep{piskunov1995, ryabchikova1997, kupka1999, kupka2000} is described by \cite{shultz_rotmag_paper1}. After cleaning, 218 spectral lines remained in the line mask. A mean Land\'e factor and mean wavelength of 1.2 and 500 nm respectively were used to extract the LSD profiles, with a velocity step of 3.6 \kms. Before the longitudinal magnetic field \bz~was measured (e.g., \citealt{mat1989}), LSD Stokes $I$ profiles were disentangled using the same procedure as described in \S~\ref{sec:rv} for the Mg~{\sc ii}~448.1 nm line, thus ensuring that the equivalent width (EW) normalization reflected only the contribution of the magnetic Aa component. Each disentangled LSD profile was shifted to the rest frame of the Aa component, and \bz~was then measured with an integration range of of $\pm 70$~\kms, and the same mean Land\'e factor and wavelength as used to extract the LSD profiles. The resulting \bz~measurements are provided in Table \ref{bztab}, along with the null \nz~measurements, and the detection flags evaluating the significance of the signal in the Stokes $V$ profile (non-detection or ND, marginal detection or MD, or definite detection or DD) according to the criteria given by \cite{1992AA...265..669D,d1997}. All but 5 Stokes $V$ profiles yield a DD, with 3 NDs. However, one of the HARPSpol $N$ profiles also yields a DD. This is likely a consequence of the RV change of the magnetic component between sub-exposures, which is a greater problem for the HARPSpol observations due to the longer sub-exposure times. The LSD profiles are shown arranged in order of orbital phase (Eqn. \ref{ephem}) in Fig.\ \ref{lsd}, where the  contribution of the B component has been minimized by removing its mean disentangled profile. The Stokes $V$ signature is clearly associated with the Aa component, as it follows the RV motion of its mean line, and is always located within the bounds of its line profile.

The \bz~and \nz~periodograms are shown in the middle panel of Fig.\ \ref{periods}. Maximum power in the \bz~periodogram is at 1.5804(26)~d, with a FAP of $10^{-4}$ in \bz~and a minimum \nz~FAP of 0.07. Assuming that the Aa component's magnetic field is, like all other magnetic fields in early-type stars, a stable fossil magnetic field, this period should correspond to the star's rotational period. This period is identical, within uncertainty, to the period recovered from the Aab RVs. \bz~is shown phased using Eqn. \ref{ephem} in the middle panel of Fig.\ \ref{bzrvew}, where the period determined from the Ab component's RVs was used due to its higher precision. In Fig.\ \ref{lsd}, Stokes $V$ also varies coherently when phased with Eqn.\ \ref{ephem}. There is no inter-cycle variability in Stokes $V$, consistent with a fossil magnetic field (Fig.\ \ref{lsd}). 

While there is a degree of uncertainty in the shape of the Aa Stokes $I$ profile due to the imperfect removal of the B component, there is clearly line profile variability that is not associated with the B component. These variations are coherent with rotational phase and are consistent between rotational cycles (see Fig.\ \ref{lsd}), consistent with an origin in surface chemical abundance inhomogeneities rather than e.g.\ pulsations. Stokes $V$ also exhibits some apparent departures from the classic S-shape expected for a purely dipolar field, particular between phases 0.9 to 0.1. However, these correspond to line profile variations in Stokes $I$, and are thus likely to be a consequence of surface abundance inhomogeneities. 

\subsection{Ab}

To constrain the magnetic field of the Ab component, we used the same LSD profiles as were used for HD 156324Aa, but measured \bz~only when the line profiles of the two stars are clearly separated. All observations yield NDs in both $N$ and Stokes $V$, with a median error bar of 1300~G and a standard deviation in \bz~of 1500~G, where the much larger uncertainties than obtained from the same LSD profiles for HD 156324Aa are an inevitable consequence of the star's much weaker lines. \bz~and \nz~are provided in Table \ref{bztab}.

\subsection{B}

\begin{figure}
\centering
\includegraphics[width=\hsize]{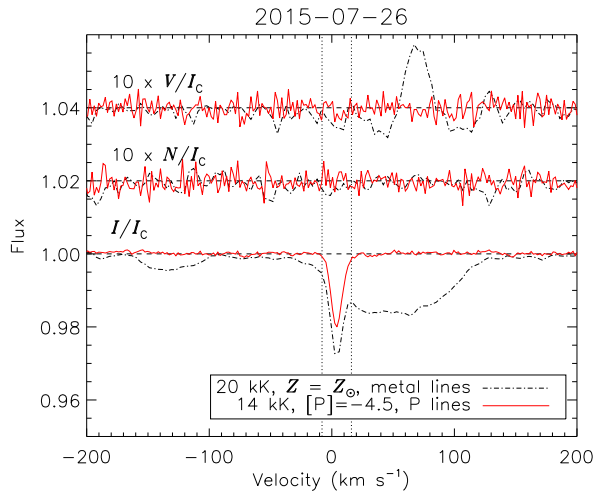}
\caption{Comparison between LSD profile extracted using a 22 kK solar metallicity mask with all metallic lines (black dot-dashed lines), and an LSD profile extracted from the same spectrum using a 14 kK mask with enhanced P, using only the P lines (red solid lines). The original LSD profile shows contributions from all 3 stars (Ab on the left, Aa on the right), while the only contribution to the Stokes $I$ profile of the P mask is from HD156324B. Vertical dotted lines show the integration range used for measuring \bz.}
\label{P_lsd}
\end{figure}

Unlike HD 156324Ab, HD 156324B is blended with HD 156324Aa in essentially all observations. However, the system's phosphorous lines are dominated by the flux from the B component due to the PGa star's extremely strong phosphorous lines. We therefore extracted LSD profiles using a 14~kK line mask (see \S~\ref{subsec:stellarpars}) with enhanced P abundances ($[{\rm P}] = -4.5$), based on the abundance analysis of the PGa star HD 19400 by \citealt{2014MNRAS.442.3604H}). The mask was cleaned so as to remove all lines obviously blended with telluric lines, interstellar lines, H lines, or lines obviously blended with HD156324Aa's spectrum, with 44 P lines remaining for analysis. As before, a mean Land\'e factor of 1.2 and a mean wavelength of 500 nm were used, however due to the very sharp spectral lines a standard velocity pixel of 1.8~\kms~was adopted. Fig.\ \ref{P_lsd} compares a representative P line LSD profile to an LSD profile from the same spectrum extracted as described in \S~\ref{sec:mag}. The P LSD profile shows essentially no contribution from HD Aa or Ab in Stokes $I$, and we can expect the same will be true in Stokes $V$. All observations yield NDs in both $N$ and Stokes $V$. The lack of evidence for a magnetic field in this star is consistent with previous failures to detect magnetic fields in PGa stars \citep{2014MNRAS.442.3604H} as well as their cooler analogues the HgMn stars \citep{2011A&A...525A..97M}. The standard deviation of \bz~is 31~G, as compared to 68~G in \nz, with a median error bar of 79~G. \bz~and \nz~are provided in Table \ref{bztab}.

\section{H$\alpha$ emission}\label{sec:halpha}

Under the assumption that HD 156324's H$\alpha$ emission is formed in the Aa component's magnetosphere, the emission variability should also be modulated with the star's rotation. To test this, we measured the equivalent width of H$\alpha$ between $\pm 0.8$~nm of the line's laboratory wavelength, after first shifting the line profiles to the reference frame of the Aa component by subtracting the RVs measured in \S~\ref{sec:rv}. The integration range was chosen to encompass only the spectral region with emission. The EWs are listed in Table \ref{rvewtab}. The H$\alpha$ EW periodogram is shown in the bottom panel of Fig.\ \ref{periods}. Maximum power is 1.5806(3)~d, identical within uncertainty to the results obtained for the RVs and \bz. The FAP is $2\times 10^{-7}$, with a null FAP of 0.51. H$\alpha$ EWs are shown phased with Eqn.\ \ref{ephem} in the bottom panel of Fig.\ \ref{bzrvew}. Since the same periods are obtained for both H$\alpha$ and \bz, the hypothesis that H$\alpha$ is formed in the star's magnetosphere seems justified. 

\begin{figure}
\centering
\includegraphics[width=\hsize]{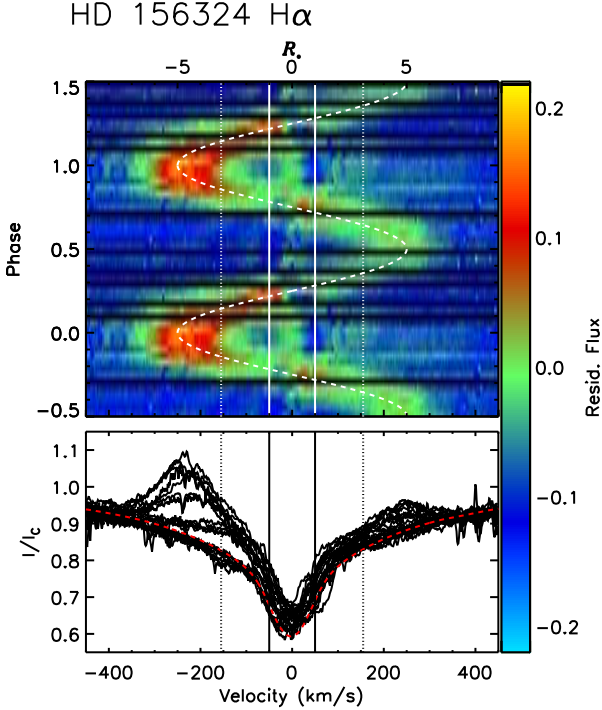}
\caption{H$\alpha$ dynamic spectrum ({\em top}) and 1D intensity ({\em bottom}). The dashed red line shows the mean synthetic profile (Aa+Ab+B) used as a reference spectrum (while this example profile did not include the RV shifts of the components, the actual synthetic line profiles used to obtain the residual flux included RV shifts). Line profiles have been moved to the Aa component's rest velocity. Solid vertical lines show \vsini; dotted lines show \rk; the dashed sinusoid shows the path of the emission profile in velocity space. Note that there is some emission at all phases, as expected given the star's small rotational inclination. Note also that there is no sign of a second cloud, which would result in a second sinusoidal emission variation in antiphase with the first.}
\label{halpha_dyn}
\end{figure}

Fig.\ \ref{halpha_dyn} shows a dynamic spectrum of H$\alpha$. Reference spectra were created from synthetic ATLAS9 spectra, using the \teff, $\log{g}$, and luminosity ratios determined for the three components by \cite{alecian2014} (see also Table \ref{phystab}). Composite spectra were created by shifting the component spectra by their measured RVs and combining them according to their luminosity ratios. The bottom panel of Fig.\ \ref{halpha_dyn} shows the observed spectra and a mean composite synthetic spectrum. Residual flux was obtained by subtracting the composite synthetic spectra from the observed spectra. In the top panel of Fig.\ \ref{halpha_dyn} the residual flux is shown phased with Eqn. \ref{ephem}, in the rest frame of the primary (i.e.\ with the Aa component's RVs subtracted). 

There is only a single emission peak in H$\alpha$, located in the blue wing at phase 0.0. While the strongest emission occurs at phase 0.0, there is detectable emission at all rotational phases. The dashed curve in Fig.\ \ref{halpha_dyn} traces its path across the line profile. Since the plasma is assumed to be in enforced corotation with the magnetic field \citep{town2005c}, radial velocity maps directly to projected radius; thus, the top horizontal axis in Fig.\ \ref{halpha_dyn} is given in units of $R_*$, with $\pm 1 R_* = \pm v\sin{i}$ indicated by vertical solid lines. At phase 0.0 the emission extends from an inner radius of approximately 2~$R_*$ to an outer radius of approximately 7.5~$R_*$, with the emission peak at about 4.75~$R_*$. Since the cloud is still in emission when its radial velocity approaches 0~\kms~at about phase 0.25, there is no evidence that it is eclipsing the star. At this phase the cloud extends between approximately $\pm 5 R_*$ projected on the plane of the sky. 

\section{Discussion}\label{sec:discussion}

\begin{table}
\centering
\caption{Orbital, physical, rotational, magnetic, and magnetospheric properties of the HD 156324 system and its components. Properties obtained from previous works are indicated with superscripts according to the following reference key: $a$) \protect\cite{alecian2014}; $b$) \protect\cite{2005AA...438.1163K}; $c$) \protect\cite{vink2001}. Upper limits for $B_{\rm d}$ for Ab and B correspond to the 99.7\% Bayesian credibility (see text).}
\label{phystab}
\resizebox{8.5 cm}{!}{
\begin{tabular}{l c c c}
\hline
\hline
Parameter & \multicolumn{3}{c}{Stellar Component} \\
          & Aa & Ab & B \\
\hline
\multicolumn{4}{c}{Orbital Parameters} \\
$P_{\rm orb}$~(d)                & \multicolumn{2}{c}{1.5805(1)} & 6.67(2) \\
${\rm JD}0$                     & \multicolumn{2}{c}{6127.29(2)} & 7208.6(3) \\
(${\rm HJD} - 2450000$)                 &  & & \\
$e$                              & \multicolumn{2}{c}{$<0.01$} & $<0.02$ \\
$v_0$~(\kms)                     & \multicolumn{2}{c}{$4 \pm 5$} & $-4.4\pm1.3$ \\
                                 &                 &             & $25 \pm 5$ \\
$K$~(\kms)                       & $51 \pm 9$ & $131 \pm 10$ & $12.6 \pm 2.1$ \\
$\omega$~($^\circ$)              & \multicolumn{2}{c}{$2.6\pm4.1$} & $<10$ \\
$a\sin{i}$~(AU)                  & \multicolumn{2}{c}{$0.026\pm 0.003$} & -- \\
$a\sin{i}$~($R_{\rm Aa}$)        & \multicolumn{2}{c}{$1.5\pm 0.2$} & -- \\
$M_{\rm tot}\sin^3{i}~(M_\odot)$ & \multicolumn{2}{c}{$0.98\pm 0.02$} & -- \\
$M\sin^3{i}~(M_\odot)$           & $0.71 \pm 0.19$ & $0.28 \pm 0.09$ & -- \\
$i_{\rm orb}~(^\circ$)           & \multicolumn{2}{c}{25.6$\pm$2.2} & -- \\
\hline
\multicolumn{4}{c}{Physical Parameters} \\
\teff~(kK)$^a$                   & 22$\pm$3 & 15.5$\pm$1.5 & 14.5$\pm$1.5 \\
$\log{g}$                        & 4.2$\pm$0.03 & 4.32$\pm$0.02 & 4.32$\pm$0.02 \\
d~(pc)$^b$                       & \multicolumn{3}{c}{1300$^{+500}_{-200}$} \\
$\log{(L/L_\odot)}$              & 3.5$\pm$0.2 & 2.4$\pm$0.2 & 2.25$\pm$0.2 \\
Age~(Myr)$^b$                    & \multicolumn{3}{c}{$7.3 \pm 3.2$} \\
$R_*~(R_\odot)$                  & 3.8$\pm$0.3 & 2.3$\pm$0.1 & 2.2$\pm$0.1 \\
$M_*~(M_\odot)$                  & 8.5$\pm$1.5 & 4.1$\pm$0.3 & 3.8$\pm$0.3 \\
$\epsilon$                       & 0.38$\pm$0.03 & & \\
\hline
\multicolumn{4}{c}{Rotational Parameters} \\
$P_{\rm rot}$~(d) (\bz)            & 1.5804(3) & -- & -- \\
$P_{\rm rot}$~(d) (H$\alpha$)      & 1.5806(3) & -- & -- \\
${\rm JD}0$ (\bz)    & 6127.3(1) & -- & -- \\
(${\rm HJD} - 2450000$) & & & \\
${\rm JD}0$ (H$\alpha$)    & 6125.54(6) & -- & -- \\
(${\rm HJD} - 2450000$) & & & \\
\vsini~(\kms)                   & 53$\pm$10 & 32$\pm$10 & 5$\pm$2 \\ 
$v_{\rm mac}$~(\kms)            & 44$\pm$29 & 31$\pm$25 & 9$\pm$3 \\ 
$W$                              & 0.18$\pm$0.02 & -- & -- \\
$v_{\rm eq}$~(\kms)              & 122$\pm$10    & -- & -- \\
$R_{\rm K}~(R_*)$                & 3.0$\pm$0.2   & -- & -- \\
$i_{\rm rot}~(^\circ)$           & 23.1$\pm$2.5 & -- & -- \\
\hline
\multicolumn{4}{c}{Magnetic Parameters} \\
$r$                              & -0.12$\pm$0.09 & -- & -- \\
$\beta~(^\circ)$                 & 71$\pm$2       & -- & -- \\
$B_{\rm d}~({\rm kG})$           & 14$\pm$1.5     & $<2.6$ & $<0.7$ \\
\hline
\multicolumn{4}{c}{Magnetospheric Parameters} \\
$\log{[\dot{M}/(M_\odot~{\rm yr}^{-1})]}^c$ & -9.0$\pm$0.9 & -- & -- \\
$v_\infty$~(\kms)$^c$                       & 1500$\pm$500 & -- & -- \\
$\eta_*$                                    & 2.0$^{+9}_{-0.8}\times 10^5$ & -- & -- \\
$R_{\rm A}~(R_*)$                                & 22$^{+11}_{-3}$ & -- & -- \\
$\tau_{\rm J}$~(Myr)                        & 2.2$^{+6.5}_{-1.5}$ & -- & -- \\
$t_{\rm S,max}~({\rm Myr})$                       & 4$^{+10}_{-3}$ & -- & -- \\
\hline
\hline
\end{tabular}
}
\end{table}

The same period was recovered from both \bz~and H$\alpha$ EWs. Phasing these measurements, as well as Stokes $V$ and H$\alpha$ line profiles, results in a coherent variation with a clear phase relationship between the different datasets in all cases. The same period was determined for the RVs of the Aab pair, indicating that the orbital and rotational periods are identical. This indicates that the Aab components exhibit orbital and rotational synchronization, indicative of tidal locking. In this section we examine the system's orbital and physical properties for further evidence of tidal locking, discuss the implications of this property, and attempt to determine if other quantities that can be derived from the orbital and rotational parameters are mutually consistent.

\subsection{Orbital parameters}\label{subsec:orbit}

Since the orbital synchronization timescale is shorter than the circularization timescale \citep{1981AA....99..126H}, if the eccentricity is zero we should also expect spin-orbit synchronization. 

Orbital parameters (eccentricity $e$, argument of periapsis $\omega$, systemic velocity $v_0$, and radial velocity semi-amplitudes $K_1$ and $K_2$) were constrained via a Monte Carlo algorithm that identifies the $\chi^2$ minimum across the orbital parameter space via comparison of synthetic to observed RV curves. This yielded $e < 0.01$, $\omega = 2.6 \pm 4.1^\circ$, $v_0 = 4 \pm 5$~\kms, $K_1 = 51 \pm 9$~\kms, and $K_2 = 131 \pm 10$~\kms. The low upper limit on the eccentricity is compatible with a circular orbit, and thus with the hypothesis that the A components are tidally locked. The orbital model and its uncertainties are indicated by the solid/dashed curves and shaded regions in the top panel of Fig.\ \ref{bzrvew}. The projected semi-major axis is $a\sin{i} = 0.026 \pm 0.003$~AU, about 1.5~$R_{\rm Aa}$. The mass ratio of the Aab pair is $M_{\rm Aa} / M_{\rm Ab} = K_{\rm Ab} / K_{\rm Aa} = 2.6 \pm 0.6$, and the total projected mass is $(M_{\rm Aa} + M_{\rm Ab})\sin^3{i} = 0.98 \pm 0.02~M_\odot$. From the HRD (using the rotating solar abundance evolutionary models of \citealt{ekstrom2012}), and restricting the luminosity of the Aa component to lie between the ZAMS and the TAMS ($\log{(L_{\rm Aa}/L_\odot)} = 3.8 \pm 0.6$), the mass of the Aa component is constrained to be $M_{\rm Aa} = 9 \pm 2~M_\odot$. The assumption that the star lies on the MS is justified on the basis of its membership in the Sco OB4 association, which is approximately 7~Myr old \citep{2005AA...438.1163K}. To satisfy the mass function, the inclination of the orbital axis from the line of sight must then be $i_{\rm orb} = 25.6 \pm 2.6^\circ$, yielding a semi-major axis of $a = 0.06 \pm 0.01$~AU.

HD 156324B shows two RV variations: a short-term variation with a period of 6.67(2)~d, and a longer-term change in $v_0$ with an unknown period, likely on the order of a year. Two hypotheses suggest themselves to explain these two timescales. The first is that the 6.67~d period corresponds to the orbit of B about the A sub-system, with the longer period corresponding to the orbit of Aa, Ab, and B about a fourth star, designated C. The second hypothesis is that the 6.67~d period is due to the presence of an undetected fourth star orbiting the B component, which we designate Bb; the longer variation then corresponds to the orbit of the centre of mass of the Bab sub-system about the Aab sub-system. In the first case, since both A and B are orbiting about C, we should expect to see the same variation in $v_0$ for A as is seen for B. As no such variation is seen, the second scenario seems more likely. 

The basic orbital parameters of the B component were determined in the same fashion as for HD 156324A. However, since the second star in the B sub-system has not yet been detected, these could not be used to obtain physical parameters for the orbit. Using the 2015 RVs, we found $K = 12.6 \pm 2.1$~\kms, $v_0 = -4.4 \pm 1.3$~\kms, $\omega < 10^\circ$, and $e < 0.02$. While nothing can be said about the semi-major axis or the projected mass, it is notable that the B sub-system, as the A sub-system, appears to be circularized. The small RV semi-amplitude suggests either that the orbital inclination is even smaller than for HD156324A, or that the mass of HD156324B's companion is much smaller than that of HD 156324B. The latter hypothesis is consistent with the absence of any evidence in the spectrum for a contribution from this star. The stellar mass of HD 156324B should be around $M_{\rm B} = 3.8 \pm 0.3~M_\odot$ (see below, \S~\ref{subsec:stellarpars}). If $i_{\rm orb,B} \sim i_{\rm orb,A}$, this would imply that the unseen Bb component should have a RV semi-amplitude of 68~\kms, implying a mass ratio $M_{\rm Ba}/M_{\rm Bb} = 5.4 \pm 0.9$, a projected mass of $M_{\rm Bb}\sin^3{i} = 0.06 \pm 0.03$~\kms, and thus a stellar mass of $M_{\rm Bb} = 0.7^{+0.8}_{-0.4}~M_\odot$. The bolometric luminosity ratio of Ba to Bb would be about 150$^{+350}_{-120}$, which is unlikely to be detectable (the luminosity ratio of Aa to Ab and B is respectively approximately 12 and 18, i.e.\ Bb would be even less luminous relative to Ba than either Ba or Ab are relative to Aa). If the orbital inclination is instead close to 90$^\circ$, then for $M_{\rm Ba}$ to lie within the range inferred from the HRD, the mass ratio would be even greater ($M_{\rm Ba}/M_{\rm Bb} \sim 13$), leading to $M_{\rm Bb} \sim 0.3~M_\odot$ and $K_2 \sim 170$~\kms. Such a low-mass companion would be even more difficult to detect.

If the HD 156324B sub-system is indeed orbiting around HD 156324A, a variation in the central velocity of HD 156324A in antiphase with that of B is expected. Assuming that the B component's undetected companion is of negligible mass, then the RV semi-amplitude of the A sub-system's centre of mass should be $v_{\rm A} = M_{\rm B,tot} / M_{\rm A,tot} v_{\rm B} \sim 5$~\kms. To check if there is evidence for such a variation, we performed Monte Carlo fits to the 2014 and 2015 HD 156324A RVs separately, finding $v_0 = 6 \pm 7$~\kms~in 2014 and $3 \pm 5$~\kms~in 2015. Thus, we find no evidence for a change in the central velocity of HD 156324A symmetrical with that of HD 156324B. However, the precision of the data, and the size of the dataset, may simply be insufficient to detect a change of the expected magnitude. Alternatively, the estimated mass of HD 156324B (which is somewhat speculative) could be higher than its true mass, in which case the mass ratio of the two sub-systems would be even greater, making any change in HD 156324A's central velocity even more difficult to detect. 

\subsection{Stellar parameters}\label{subsec:stellarpars}

\begin{figure}
\centering
\includegraphics[width=\hsize]{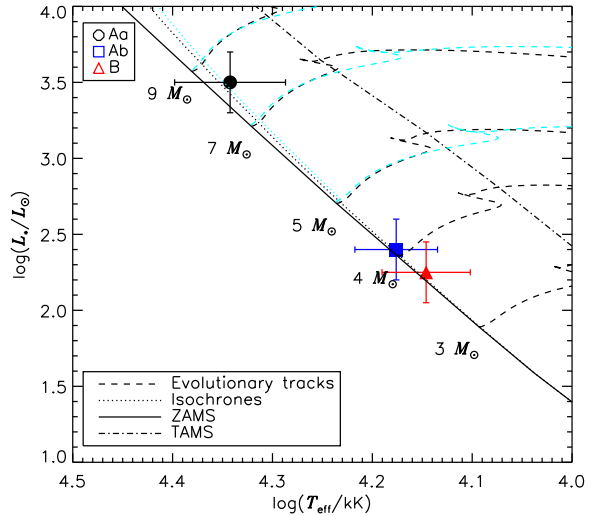}
\caption{Hertzsprung-Russell Diagram (HRD). Luminosities for Aa and Ab were determined via parallax and photometry, and were further constrained by requiring that they be consistent wth both the star's orbital parameters and with the age of the Sco OB4 association. Evolutionary tracks and isochrones are from \protect\cite{ekstrom2012} (black) and \protect\cite{2011A&A...530A.115B} (light blue), where for the former we have used their rotating models, and for the latter we used Galactic metallicity models with an initial rotational velocity of $\sim$230~\kms.}
\label{hrd}
\end{figure}

Effective temperatures were taken from the spectral modelling performed by \cite{alecian2014}, who found \teff$_{\rm , Aa} = 22 \pm 3$~kK, and \teff$_{\rm , Ab}$ and \teff$_{, {\rm B}} \sim 15.5 \pm 1.5$~kK. Since their modelling yielded the result that Ab is $\sim$1 kK hotter than B, we adopt \teff$_{, {\rm Ab}} = 15.5 \pm 1.5$~kK and \teff$_{, {\rm B}} = 14.5 \pm 1.5$~kK. 

We calculated the luminosity of HD156324Aa from the distance to the Sco OB4 association ($d=1500^{+500}_{-300}$~pc, \citealt{2005AA...438.1163K}), the extinction $A_V = 0.92 \pm 0.25$ \citep{petit2013}, and the apparent magnitude $V = 8.76$, using a bolometric correction $BC = -2.15 \pm 0.08$ obtained from the {\sc tlusty} BSTAR2006 grid \citep{lanzhubeny2007} via linear interpolation for the Aa component's \teff~and surface gravity. We assumed that $\log{g} = 4.0$ in this step. We also assumed that Ab and B contribute negligibly to the total luminosity. This yielded $\log{(L_{\rm Aa}/L_\odot)} =  3.8 \pm 0.6$. We then refined this estimate by requiring that $\log{L_{\rm Aa}}$ be consistent with the OB association age, $t = 7 \pm 3$~Myr, from which we obtained $\log{(L_{\rm Aa}/L_\odot)} = 3.5 \pm 0.2$ (see Fig.\ \ref{hrd}). From this, the radius is $R_{\rm Aa} = \sqrt{(L_{\rm Aa}/L_\odot)/(T_{\rm eff,Aa}/T_{\rm eff,\odot})^4} = 3.8 \pm 0.3~R_\odot$, where $T_{\rm eff,\odot} = 5.78$~kK. The mass, from the HRD, is $M_{\rm Aa} = 8.5 \pm 1.5~M_\odot$, which is slightly more precise than that obtained without applying the association age restriction to the luminosity. Applying this mass to the mass function in \S~\ref{subsec:orbit} yields $i_{\rm orb} = 26 \pm 2.2^\circ$.

The luminosity of Ab was first inferred from its projected mass and orbital inclination, which requires that $M_{\rm Ab} = 3.3 \pm 1.3~M_\odot$, implying $\log{(L_{\rm Ab}/L_\odot)} = 2.0 \pm 0.9$. Requiring that $\log{(L_{\rm Ab}/L_\odot)}$ be consistent with \teff$_{\rm , Ab}=15.5 \pm 1.5$~kK narrows the range to $\log{(L_{\rm Ab}/L_\odot)} = 2.6 \pm 0.3$. Further requiring that the luminosity also be consistent with the association age then yields $\log{(L_{\rm Ab}/L_\odot)} = 2.4 \pm 0.1$. The mass is then $M_{\rm Ab} = 4.1 \pm 0.3~M_\odot$. 

The timescale for spin-orbit alignment should be comparable to the timescale for synchronization, both of which are much less than the timescale for circularization \citep{1981AA....99..126H}. Since the system exhibits strong evidence of synchronization, and the RVs are compatible with $e = 0$, it is reasonable to assume that the rotational axes of the two stars must be aligned. In this case we expect that $R_{\rm Aa}/R_{\rm Ab} = v\sin{i}_{\rm Aa}/v\sin{i}_{\rm Ab}$. The ratio of stellar radii is $1.65\pm0.20$, and the ratio of measured \vsini~is $1.65\pm 0.83$, as expected. 

For HD 156324B it is not possible to use its orbital properties to constrain its luminosity, since the stellar properties of its hypothetical companion are highly uncertain. However, the spectral modelling performed by \cite{alecian2014} was consistent with the Ab and B components having essentially identical luminosities. Therefore, we started with the same luminosity range as for HD 156324Ab, and then refined it by requiring that it be consistent with \teff$_{\rm , B}$~and the cluster age, thus yielding $\log{L} = 2.25 \pm 0.2$.

\cite{alecian2014} fixed $\log{g} = 4.0$, as due to the blending of the components the surface gravity could not be determined with precision. However, from Geneva photometry they found $\log{g}=4.38$. The values given in Table \ref{phystab} are those inferred from the HRD, and not direct measurements. They are consistent with the higher surface gravity found via Geneva photometry \citep{alecian2014}, which would be expected for stars close to the ZAMS. 

Since we have used stellar evolutionary models to help constrain the luminosities and radii of the stars, it is worth comparing results obtained for different grids of evolutionary models. \cite{2011A&A...530A.115B} computed a dense grid of models for different metallicities and initial rotational velocities, and used a different core overshooting parameter from that adopted by \cite{ekstrom2012}. In Fig.\ \ref{hrd} the \cite{2011A&A...530A.115B} evolutionary tracks and isochrones are shown for comparison, where we adopted the models with Galactic metallicity and initial rotation velocities of $\sim$230~\kms. The \cite{2011A&A...530A.115B} models yield a TAMS at a lower \teff~than in the \cite{ekstrom2012} models (see Fig.\ \ref{hrd}). However, there is essentially no difference in the position of the $\log{(t/{\rm Myr})} = 6.8$ isochrone. \cite{2016ApJ...823..102C} provided a grid of evolutionary tracks and isochrones computed with MESA. Their comparison to the Geneva models calculated by \cite{ekstrom2012} (see their Fig.\ 19) shows that, for MS stars in the mass range of interest for this system, the evolutionary tracks are essentially indistinguishable. 

The impact of tidal interactions on the evolution of massive stars has been explored by \cite{2013A&A...556A.100S,2016A&A...585A.120S}. They investigated scenarios across a range of stellar masses, initial rotation velocities, orbital periods, and with and without solid body rotation imposed by internal magnetic fields. In general, these factors can have quite a profound impact on the trajectory of a star across the HRD, affecting not just the rate at which it evolves, but also the direction, e.g.\ causing a star to evolve towards higher luminosities and effective temperatures. Song et al.\ found the differences between single and binary star evolutionary models close to the ZAMS to be negligible. Since HD 156324 is a young system, it seems unlikely that modifications to its internal structure due to tidal interactions have yet had time to profoundly affect its location on the HRD. However, tidal effects should certainly not be neglected in any attempt to model the future evolution of HD 156324A. While the Song et al.\ models concern stars of much higher primary masses ($M_1 \ge 15 M_\odot$), they also considered a less extreme mass ratio ($M_1/M_2 = 1.5$, as compared to 2.6 in the present case), thus tidal interactions are likely to have a very significant impact on the evolution of HD 156324Ab, especially. 

\subsection{Rotation and Magnetic Field}

\subsubsection{Aa}

The \bz~curve in Fig.\ \ref{bzrvew} is consistent with a sinusoidal variation (the reduced $\chi^2$ of the least-squares sinusoidal fit to \bz~is 1.3), indicating that the main contribution of the surface magnetic field to \bz~is from the dipolar component, as is usually the case for magnetic early-type stars. We therefore constrain the surface magnetic field of the Aa component using the simplest possible description, a dipolar Oblique Rotator Model (ORM). A dipolar ORM consists of five parameters: the epoch JD0 of maximum \bz, the rotational period $P_{\rm rot}$, an inclination $i_{\rm rot}$ of the rotational axis from the line of sight, the obliquity angle $\beta$ between the magnetic pole and the rotational axis, and the surface strength of the magnetic field at the magnetic pole $B_{\rm d}$. 

Typically $i_{\rm rot}$ is constrained via the rotational period, \vsini$_{\rm rot}$, and the stellar radius. The large formal uncertainties in $\log{(L_{\rm Aa}/L_\odot)}$ lead to a highly uncertain radius, $R_{\rm Aa} = 5 \pm 2~R_\odot$, from which we obtain $i_{\rm rot} = 12^{\circ+23}_{-5}$. If the more precise age-restricted luminosity is used instead, the higher precision in the radius leads to $i_{\rm rot} = 23 \pm 2^\circ$. This is identical, within uncertainty, to  $i_{\rm orb}$, as would be expected for a tidally locked system. This serves as additional confirmation that the isochrones have accurately constrained the luminosity.

To solve for $\beta$ we use the relation developed by \cite{preston1967}: $\tan{\beta} = (1 - r)/(1 + r)\cot{i_{\rm rot}}$, where $r=-0.12 \pm 0.09$ is given by the ratio $r = (|B_0| - B_1)/(|B_0 + B_1|)$, and $B_0$ and $B_1$ are the mean and semi-amplitude of the least-squares sinusoidal fit shown in Fig.\ \ref{bzrvew}. This yields $\beta = 70 \pm 2^\circ$. We then obtain $B_{\rm d} = 14 \pm 1.5$~kG using Eqn.\ 1 from \cite{preston1967}, which also requires the limb darkening coefficient $\epsilon = 0.38 \pm 0.03$, which we obtained from the tables published by \cite{diazcordoves1995} using the star's \teff, and the surface gravity inferred from its age-restricted position on the HRD, $\log{g} = 4.22 \pm 0.03$. 

\subsubsection{Ab}

Since Aa is tidally locked, the less massive Ab component must be as well since it has a smaller angular momentum reservoir. Therefore its rotational period should also be 1.58~d, and individual LSD profiles can be combined to obtain higher precision. We binned LSD profiles by phase between phases 0.9 and 0.1 (7 observations) and between phases 0.4 and 0.6 (4 observations), at which phases the line profiles of the Aa and Ab components are clearly separated. Both of the mean LSD profiles yielded NDs, with \bz~$=-246 \pm 303$~G and $34 \pm 573$~G, respectively. To derive upper limits on $B_{\rm d}$, we utilized a modified version of the Bayesian analysis engine {\sc raven} described by \cite{petit2012a}. Synthetic Stokes $V$ profiles with the same line broadening parameters as the observed line profiles, and normalized to the observed Stokes $I$ LSD EW, were compared to the observed LSD Stokes $V$ LSD profiles. {\sc crow}, the modified version of {\sc raven}, phases the observations according to rotational phase, rather than treating phase as a random variable. Since the orbital and spin axes of the system are aligned, we fixed $i_{\rm rot}=i_{\rm orb}$. Since we know $P_{\rm rot}$ but not JD0 (i.e., the time at which $|\langle B_z \rangle| = |\langle B_z \rangle|_{\rm max}$), rotational phases were fixed by keeping the two observations separated by $\Delta\phi = 0.5$. The resulting upper limits, at credibility intervals of 68.3\%, 95.4\%, and 99.7\%, are respectively 314 G, 969 G, and 2.6 kG. 

\subsubsection{B}

Since neither $i_{\rm rot}$ nor $P_{\rm rot}$ are known for the B component, Bayesian analysis of the Stokes $V$ profiles was conducted using {\sc raven} \citep{petit2012a}, i.e. with $i_{\rm rot}$ drawn from a $\sin{i}$ distribution and $\phi$ from a flat distribution. The upper limits at credibility intervals of 68.3\%, 95.4\%, and 99.7\%, are respectively 72 G, 204 G, and 704 G. The very tight upper limits are a consequence of both the star's sharp spectral lines and the large number of observations, which make it unlikely that the star was observed at a rotational phase at which a relatively strong magnetic field would not have been detected due to its sky-projected geometry. 

\subsection{Magnetosphere}\label{subsec:magnetosphere}

The magnetospheres of early-type stars can be classified as either dynamical or centrifugal, depending on the role played by rotation in preventing gravitational infall of the corotating, magnetically confined plasma \citep{petit2013}. Whether or not the wind is magnetically confined in the first place can be determined by the wind magnetic confinement parameter $\eta_*$ \citep{ud2002}, which is the ratio of magnetic to kinetic energy density in the wind at the magnetic equator. From the star's age-restricted position on the HRD, and using the \cite{vink2001} mass-loss recipe, the mass-loss rate and wind terminal velocity are, respectively, $\log{[\dot{M}/(M_\odot~{\rm yr}^{-1})]} = -9.0 \pm 0.9$ and $v_\infty = 1700 \pm 500$~\kms. Using Eqn.\ 7 from \cite{ud2002}, and $B_{\rm d}$ and $R_{\rm Aa}$, we obtain $\eta_* = 2.0^{+9}_{-0.8}\times 10^4 \gg 1$, therefore the wind is strongly magnetically confined. At the magnetic equator, magnetic confinement should persist out to the Alfv\'en radius, beyond which the wind opens the magnetic field lines; this is given by Eqn.\ 7 from \cite{ud2008}, from which we obtain $R_{\rm A} = 22^{+11}_{-3} R_*$. 

The distance from the stellar surface beyond which centrifugal force overpowers gravitational force is given by the Kepler radius $R_{\rm K}$ (Eqn.\ 12, \citealt{town2005c}). Using $P_{\rm rot} = 1.58~{\rm d}$ and $M_{\rm Aa}$ and $R_{\rm Aa}$ from the age-restricted HRD position, we find $R_{\rm K} = 3.0 \pm 0.2 R_*$. Since $R_{\rm K} \ll R_{\rm A}$, HD 156324Aa has a centrifugal magnetosphere (CM). 

When emission is present, CMs invariably show a distinctive emission pattern, with two emission peaks at high projected velocities at maximum emission, which oscillate in an approximately sinusoidal fashion between the blue- and red-shifted halves of the line profile throughout a rotational cycle (e.g.\ \citealt{leone2010, bohl2011, grun2012, oks2012, rivi2013}). The high velocity emission peaks arise due to the accumulation of plasma above $R_{\rm K}$, while the pattern of variability is due to the changing projection of the plasma on the sky as the star rotates \citep{town2005c}. Since the plasma will accumulate in the gravitocentrifugal minima along magnetic field lines, the highest concentrations of plasma (`clouds') should be at the intersections of the magnetic and rotational equators; thus, if $\beta$ is large, there should be two distinct clouds; if $\beta = 0$, the plasma should be distributed in a uniform disk; and for intermediate values of $\beta$, the plasma is expected to be distributed in a warped disk \citep{town2005c}. Assuming the clouds are optically thick, emission strength is sensitive to the projected area of the magnetosphere, and should thus correlate to \bz, with maximum emission occuring at \bz$_{\rm max}$. If only a single magnetic pole is visible throughout a rotational cycle, the EW curve of an emission line formed in the CM should have only a single maximum (a single-wave variation); if two magnetic poles are visible throughout a rotational cycle, there should be two local EW maxima corresponding to the extrema of the \bz~curve (a double-wave variation). Depending on $i$ and $\beta$, it is also possible for the plasma to pass across the line of sight to the star, causing a sharp increase in absorption in the line core due to eclipsing of the star by the thick cloud.

The EW curve in Fig.\ \ref{bzrvew} correlates to the \bz~curve in the manner expected, with maximum emission close to \bz$_{\rm max}$ (with an offset of 0.1 cycles), and a single-wave variation consistent with a \bz~curve that shows only a single magnetic pole. Examining the dynamic spectrum in Fig.\ \ref{halpha_dyn}, HD 156324Aa shows no sign of enhanced absorption in the line core, which is consistent with its small inclination, at which no eclipsing should take place \citep{town2008}. At maximum emission (phase 0.0), the emission is predominantly located above $R_{\rm K}$ (indicated by the vertical dotted lines in Fig.\ \ref{halpha_dyn}). However, there is evidence of only a single cloud, a marked departure from the usual pattern for stars with CMs. This is clear from Fig.\ \ref{halpha_dyn}: the emission follows a single sinusoidal path, as indicated by the dashed curve, with no sign of another emission bump varying in antiphase as would be expected if there were two clouds. 

One possibility to explain this anomolous emission pattern could be a complex surface magnetic field with strong contributions from non-dipolar components. For instance, $\sigma$ Ori E's surface magnetic field has a significant quadrupolar contribution as revealed by Zeeman Doppler Imaging, and extrapolating this topology into the circumstellar environment yields an asymmetrical CM which is able to reproduce its asymmetrical emission structure \citep{2015MNRAS.451.2015O}. However, $\sigma$ Ori E still displays two distinct emission bumps. Furthermore, there is no evidence for anharmonicity in HD 156324Aa's \bz~curve (Fig.\ \ref{bzrvew}). 

\begin{figure}
\centering
\includegraphics[width=\hsize]{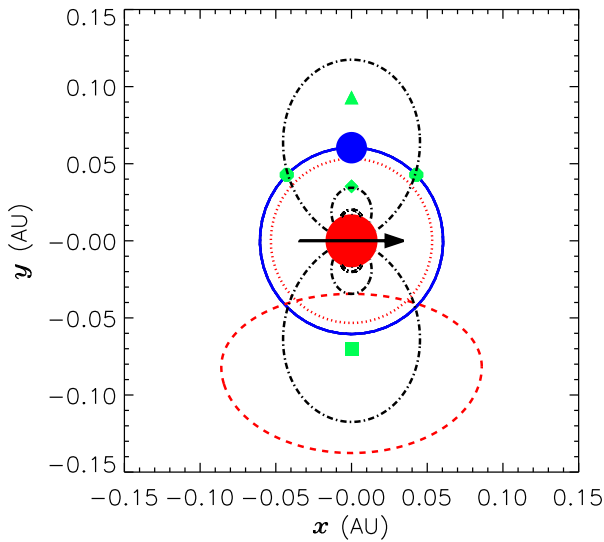}
\caption{Scale schematic of the HD 156324A system in the rest frame of the Aa component. The filled circles indicate Aa (red) and Ab (blue). The solid (blue) line shows the orbit of Ab. The dotted (red) line indicates \rk. The dashed (red) line indicates the estimated extent of the H$\alpha$ emission. The green points indicate the Lagrange points: ($L_1$, $L_2$, $L_3$, $L_4$, and $L_5$, respectively diamond, triangle, square, and circles). Dot-dashed (black) lines show representative magnetic field lines, and the magnetic axis is indicated by the arrow. \ra~extends out to $\sim$0.35 AU, and is not shown on the plot i.e. the entire system can be considered magnetically confined. Note that \rk~and the semi-major axis of the system are, within uncertainties, essentially identical; also note that Ab occupies the expected position of the missing magnetospheric cloud.}
\label{magbinorb}
\end{figure}

The unique orbital properties of the system offer another possible solution. Since a CM is a consequence of magnetic confinement within a gravitocentrifugal potential, modification of this potential should change the accumulation surface and, hence, the equilibrium distribution of circumstellar plasma. Fig.\ \ref{magbinorb} shows a scale schematic of the HD156324A system. The semi-major axis and \rk~are, within their uncertainties, essentially identical (as would be expected for a tidally locked system). The approximate extent of the cloud, as determined via visual inspection of the H$\alpha$ dynamic spectrum (see Fig.\ \ref{halpha_dyn} and \S~\ref{sec:halpha}), is indicated on the plot. From comparing the RV and EW curves in Fig.\ \ref{bzrvew}, the centre of the cloud should be approximately opposite to the position of Ab. Thus, Ab occupies the expected place of the missing cloud. Furthermore, since \bz$_{\rm max}$ and RV$_{\rm max}$ approximately coincide, we can infer that the magnetic axis is approximately tangential to the orbit (i.e. perpendicular to the line connecting Aa and Ab), as indicated in Fig.\ \ref{magbinorb}. As noted previously, there is a small phase offset of about 0.1 cycles between \bz$_{\rm max}$ and maximum emission. This could point towards either a second-order modification of the gravitocentrifugal potential, or to slight departures from a purely dipolar magnetic field. 

The Lagrange points, indicated on the plot, provide an indication of the locations of the potential minima of the system. $L_4$ and $L_5$ are dynamically unstable, and are furthmore not in the magnetic equator, thus material flowing along magnetic loops intersecting these points will be free to move along magnetic field lines and should not accumulate. 

$L_1$ through $L_3$ are all saddle points, dynamically unstable along the line connecting the stars, and stable along the direction of the orbit. All three points lie in the magnetic equator. Assuming that the ratio of magnetic energy density to wind kinetic energy density is high enough that the plasma is locked to the magnetic field, as in the RRM model, the magnetic field will stabilize perturbations along the the line connecting the two stars i.e.\ all three points are dynamically stable. However, there is no evidence for any significant amount of plasma anywhere but at $L_3$, which is located just above \rk~(and, indeed, is close to the radius of maximum emission at about 4.75 $R_{\rm Aa}$). The absence of emission at $L_2$ is likely because this is opposite Ab, which may block or disrupt mass loading via Aa's wind. $L_1$ is located below \rk, thus centrifugal force at this point is insufficient to prevent gravitational infall.

\subsection{Magnetic Braking}

Magnetic stars are expected to rapidly lose angular momentum via their magnetospheres \citep{wd1967,ud2009}. With the mass-loss rate and Alfv\'en radius from \S~\ref{subsec:magnetosphere}, and with the radius of gyration $r_{\rm gyr} = 0.25$ from the stellar structure models of \cite{claret2004}, the spindown timescale is $\tau_{\rm J} = 1.7^{+3.6}_{-1.2}$~Myr. The maximum spindown age, obtained under the assumption of intially critical rotation, is then $t_{\rm S,max} = 3^{+6}_{-2}$~Myr, which is compatible with the age of the cluster, $7 \pm 3$~Myr. This spindown rate corresponds to a change in the rotational period of $\dot{P} = 80^{+270}_{-26}~{\rm ms~yr^{-1}}$. While this is much less than the current precision in the period of about 9~s, it should in principle be detectable with a larger spectroscopic or photometric dataset since, as the temporal baseline of the measurements increases, so does the precision of the period. Period changes of this magnitude have been measured for a few stars (HD 37776; \citealt{miku2008,miku2011}; CU VIR, \citealt{miku2008}; $\sigma$ Ori E, \citealt{town2010}), for which the datasets span about 30 years. 

Since the magnetic star is in a tidally locked binary system, it is probable that its angular momentum evolution will be affected by its orbital properties. The interaction of magnetic braking and tides on the evolution of a close binary's rotation has been examined by \cite{2017arXiv170901902S}. They argued that magnetic braking will always act to spin the star down, while tides will either accelerate or decelerate a star's surface rotation depending on whether the angular frequency of rotation is less than or greater than the orbital angular frequency. \citeauthor{2017arXiv170901902S} examined two scenarios, $P_{\rm orb} > P_{\rm rot}$ and $P_{\rm orb} < P_{\rm rot}$. In the former case, both tides and magnetic braking initially work together to slow the star's rotation, resulting in rapid spindown. In the latter, the timescale $\tau_{\rm tide}$ over which tides transfer angular momentum from the orbit to the star's rotation are much shorter than the magnetic braking timescale even for very strongly magnetized stars; thus the star's rotation is initially accelerated by tidal interaction. In either case, once the spin and orbital periods are synchronized, $\tau_{\rm tide} \sim \tau_{\rm J}$, and the star settles into an equilibrium rotational frequency slightly below the orbital frequency. Since the equilibrium rotational angular frequency is always less than the orbital angular frequency, an important difference between these two scenarios is that there is a time at which $P_{\rm rot} \sim P_{\rm orb}$ only in the spin-down scenario, as in the spin-up scenario $P_{\rm rot}$ will be less than $P_{\rm orb}$ at all times. 

Since HD 156324 has essentially identical orbital and rotational periods (to within 0.0001~d, approximately the same as the precision of $P_{\rm orb}$), only the spin-down scenario can apply. Furthermore, since the timescale over which $P_{\rm rot} = P_{\rm orb}$ is predicted to be quite short, the model of \citeauthor{2017arXiv170901902S} implies that we must be observing this system during a brief window in its evolution. Future observations should aim to refine the precision of $P_{\rm rot}$ and $P_{\rm orb}$, and to determine if they are in fact exactly identical. Determining $\dot{P_{\rm rot}}$ and $\dot{P_{\rm orb}}$ will also be a useful test: the model proposed by \cite{2017arXiv170901902S} would predict that $P_{\rm rot}$ should slow at about the rate predicted by $\tau_{\rm J}$, while $P_{\rm orb}$ should remain constant. This should continue for the forseeable future. Using Eqn.\ 3~from \cite{2017arXiv170901902S}, which gives the ratio $\tau_{\rm tide} / \tau_{\rm J}$, and using the orbital, rotational, physical, and magnetic parameters in Table \ref{phystab}, we see that $\tau_{\rm J} > \tau_{\rm tide}$ until $P_{\rm rot} \sim 4.3~{\rm d}$, i.e.\ the star should continue slowing down for about another 3~Myr. 

\cite{2017arXiv170901902S} also predicted that the interplay between magnetic braking and tidal interactions should have a profound influence on stellar evolution, and that this is particularly the case for the spin-down scenario (see their Fig.\ 5). The details of these effects depend sensitively on whether internal differential or solid-body rotation is assumed; while there are theoretical reasons to expect magnetic stars to rotate as solid bodies, it is not yet known if this is the case. However, across all of the models considered by \cite{2017arXiv170901902S}, the differences in evolutionary models close to the ZAMS are less than $0.02$ dex in $\log{T_{\rm eff}}$ and less than 0.1 dex in $\log{L}$; since these are smaller than the uncertainties of 0.05 dex in $\log{T_{\rm eff}}$ and 0.2 dex in $\log{L}$ for HD 156324Aa, the fundamental parameters derived in \S~\ref{subsec:stellarpars} should not be affected by neglecting the impact of magnetic braking on stellar evolution. 

\subsection{Magnetic interactions}\label{subsec:maginter}

The alignment of Aa's magnetic dipole in the orbital plane, approximately perpendicular to the line connecting the two stars, does not seem like a random configuration, but rather appears to be a potential energy minimum. While there is no evidence for a magnetic field in Ab, a  kG magnetic field could remain undetected. Two possible cases should thus be evaluated: one in which Ab is magnetic, and one in which Ab is not magnetic.

Only one example is known of a doubly magnetic hot binary, the B2/B3 system $\epsilon$ Lupi \citep{2015MNRAS.454L...1S}. In this system, both stars are of approximately similar masses, in an eccentric 4.6~d orbit. Their weak ($\le 1~{\rm kG}$) magnetic dipoles are approximately anti-aligned with one another, and hence (since the rotational and orbital axes are likely to be aligned in this system) also aligned with the orbital axis. This is a very different configuration from that exhibited by HD 156324. This points either to Ab being non-magnetic, or to the existence of multiple potential energy minima towards which magnetic binaries can evolve. 

Assuming that Aa's magnetic dipole falls in strength as $1/r^2$, and given that the stars are separated by about $3.4~R_{\rm Aa}$ (Table \ref{phystab}), the magnetic field at Ab should be about 350~G, i.e.\ well within the range observed for magnetic B-type stars in general. If Ab is a non-magnetic star, it is interesting to ask if the close proximity of the two stars might result in a magnetic field being induced within Ab. During the system's evolution, prior to tidal locking, the magnetic field lines of Aa would have been in continuous motion relative to Ab. The relative motion of a magnetic field and a large body of plasma would then lead to induced electrical currents, likely resulting in continous magnetic reconnection. This would lead to some quantity of ions being stripped off of the outer layers of Ab's photosphere, possibly slowing the rotation of both stars and, thus, accelerating tidal locking. 

Based upon 3D MHD simulations performed by \cite{2016ApJ...833..140S} and \cite{2015ApJ...815..111S}, \cite{2017ApJ...847L..16S} showed that, for cool stars and closely orbiting exoplanets, magnetic torques can under certain circumstances dominate over tidal torques, thus decreasing the migration timescale. This is particularly the case when the primary is strongly magnetized. Since their calculations were for stars with convective envelopes orbited by either hot Jupiters or rocky super-Earths, they cannot necessarily be extrapolated directly to the case of two stars of similar masses with radiative envelopes. However, given the close proximity of the Aab pair, and that the orbital system is well within the Alfv\'en sphere of the magnetic star, their results raise the intriguing possibility that the magnetic torque could contribute significantly or even primarily to the orbital evolution of the system, and could moreover lead to a rapid inward migration of the secondary star. 

\section{Conclusions}\label{sec:conclusions}

We have determined the orbital periods $P_{\rm orb}$ of HD 156324A and B, and the rotational period $P_{\rm rot}$ of HD 156324Aa, the latter through both magnetic and spectroscopic diagnostics. For HD 156324Aa, $P_{\rm orb} = P_{\rm rot}$, indicating that the system is tidally locked. In keeping with this, the orbit has been circularized. The semi-major axis of the system, determined via orbital modelling, and the Kepler cortation radius, determined from the star's stellar and rotational properties, are identical within error bars, as expected. We furthermore obtain similar values for the orbital and rotational inclinations via independent methods, thus demonstrating that the orbital and rotational axes are aligned, as must be the case for a tidally locked system. 

HD 156324Aa has a strong magnetic field ($B_{\rm d} \sim 14$~kG), with a large tilt between its rotational and magnetic axes ($\beta \sim 70^\circ$). No magnetic field was detected in either the Ab or B components, with upper limits of $B_{\rm d} \le 2.6$~kG for Ab and $0.7$~kG for B determined via Bayesian analysis of their LSD Stokes $V$ profiles. The results for B are in agreement with the analysis of another PGa star by \cite{2014MNRAS.442.3604H}, supporting that this class of stars, like their cooler analogues the HgMn stars, do not generally possess magnetic fields stronger than $\sim 10-100$~G.

In keeping with its relatively rapid rotation, strong magnetic field, and early spectral type, HD 156324 Aa displays H$\alpha$ emission which both looks and behaves approximately as expected for emission originating in a centrifugal magnetosphere. However, in sharp contrast with the CMs seen around single stars, HD 156324Aa shows evidence of only one magnetospheric cloud. It seems very likely that this is a consequence of orbital disruption of the gravitocentrifugal potential, since HD 156324Ab occupies the expected place of the missing cloud. 

The central velocity of HD 156324B exhibits a long term variation that is probably a consequence of the orbit of its system's centre of mass about that of HD 156324A. While no symmetrical change in the central velocity of HD 156324A could be detected, this might be due to a lower than estimated mass for HD 156324B, or data of insufficient precision. Further observation will be necessary to determine the period and amplitude of HD 156324B's long term variation.

This remarkable system, with its unique combination of rotational, magnetic, and orbital phenomena, will be an important target for investigation of the consequences of orbital effects on CMs. In principle, it should be amenable to a relatively simple modification of the semi-analytic, time-independent RRM model \citep{town2005c}. Its short period and rapid spindown timescale means that rotational spindown should be detectable in the future. This may prove important for investigating what role, if any, magnetic fields play in regulating the change of the system's orbital properties. 

\section*{Acknowledgements} This work has made use of the VALD database, operated at Uppsala University, the Institute of Astronomy RAS in Moscow, and the University of Vienna. This work is based on observations obtained at the Canada-France-Hawaii Telescope (CFHT) which is operated by the National Research Council of Canada, the Institut National des Sciences de l'Univers of the Centre National de la Recherche Scientifique of France, and the University of Hawaii. All authors acknowledge the advice and assistance provided on this and related projects by the members of the BinaMIcS and MiMeS collaborations. MS acknowledges the financial support provided by the European Southern Observatory studentship program in Santiago, Chile. MS and GAW acknowledge support from the Natural Sciences and Engineering Research Council of Canada (NSERC). EA acknowledges financial support from the Programme National de Physique Stellaire (PNPS) of INSU/CNRS. We acknowledge the Canadian Astronomy Data Centre (CADC). This project used the facilities of SIMBAD.

\bibliography{bib_dat.bib}{}

\end{document}